\documentclass[twocolumn]{aastex631}
\usepackage[utf8]{inputenc}
\usepackage{amsmath}
\usepackage{bbold}
\usepackage{natbib}

\def\Ha{\hyperlink{cite.h1c_idr2_limits}{H22a}}
\def\Hb{\hyperlink{cite.h1c_idr2_theory}{H22b}}
\def\Hthree{\hyperlink{cite.h1c_idr3_limits}{H23}}

\begin{document}

\title{Exploring One-point Statistics in HERA Phase I Data: Effects of Foregrounds and Systematics on Measuring One-Point Statistics}

\correspondingauthor{Honggeun Kim}
\email{hgkim@mit.edu}

\author[0000-0001-5421-8927]{Honggeun  Kim}
\affiliation{MIT Kavli Institute, Massachusetts Institute of Technology, Cambridge, MA}
\affiliation{Department of Physics, Massachusetts Institute of Technology, Cambridge, MA}

\author[0000-0002-4117-570X]{Jacqueline N. Hewitt}
\affiliation{MIT Kavli Institute, Massachusetts Institute of Technology, Cambridge, MA}
\affiliation{Department of Physics, Massachusetts Institute of Technology, Cambridge, MA}

\author[0000-0002-8211-1892]{Nicholas S. Kern}
\affiliation{MIT Kavli Institute, Massachusetts Institute of Technology, Cambridge, MA}
\affiliation{Department of Physics, Massachusetts Institute of Technology, Cambridge, MA}
\affiliation{NASA Hubble Fellow}

\author[0000-0003-3336-9958]{Joshua S. Dillon}
\affiliation{Department of Astronomy, University of California, Berkeley, CA}

\author[0000-0002-3839-0230]{Kai-Feng Chen}
\affiliation{MIT Kavli Institute, Massachusetts Institute of Technology, Cambridge, MA}
\affiliation{Department of Physics, Massachusetts Institute of Technology, Cambridge, MA}

\author[0000-0001-5112-2567]{Zhilei  Xu}
\affiliation{MIT Kavli Institute, Massachusetts Institute of Technology, Cambridge, MA}

\author{Eleanor  Rath}
\affiliation{MIT Kavli Institute, Massachusetts Institute of Technology, Cambridge, MA}
\affiliation{Department of Physics, Massachusetts Institute of Technology, Cambridge, MA}

\author[0000-0002-3317-5347]{Vincent MacKay}
\affiliation{MIT Kavli Institute, Massachusetts Institute of Technology, Cambridge, MA}

\author{Tyrone  Adams}
\affiliation{South African Radio Astronomy Observatory, Black River Park, 2 Fir Street, Observatory, Cape Town, 7925, South Africa}

\author[0000-0002-4810-666X]{James E. Aguirre}
\affiliation{Department of Physics and Astronomy, University of Pennsylvania, Philadelphia, PA}

\author{Rushelle  Baartman}
\affiliation{South African Radio Astronomy Observatory, Black River Park, 2 Fir Street, Observatory, Cape Town, 7925, South Africa}

\author[0000-0001-9428-8233]{Adam P. Beardsley}
\affiliation{School of Earth and Space Exploration, Arizona State University, Tempe, AZ}
\affiliation{Department of Physics, Winona State University, Winona, MN}

\author[0000-0002-0916-7443]{Gianni  Bernardi}
\affiliation{INAF-Istituto di Radioastronomia, via Gobetti 101, 40129 Bologna, Italy}
\affiliation{Department of Physics and Electronics, Rhodes University, PO Box 94, Grahamstown, 6140, South Africa}
\affiliation{South African Radio Astronomy Observatory, Black River Park, 2 Fir Street, Observatory, Cape Town, 7925, South Africa}

\author{Tashalee S. Billings}
\affiliation{Department of Physics and Astronomy, University of Pennsylvania, Philadelphia, PA}

\author[0000-0002-8475-2036]{Judd D. Bowman}
\affiliation{School of Earth and Space Exploration, Arizona State University, Tempe, AZ}

\author{Richard F. Bradley}
\affiliation{National Radio Astronomy Observatory, Charlottesville, VA}

\author[0000-0001-5668-3101]{Philip  Bull}
\affiliation{Jodrell Bank Centre for Astrophysics, University of Manchester, Manchester, M13 9PL, United Kingdom}
\affiliation{Department of Physics and Astronomy,  University of Western Cape, Cape Town, 7535, South Africa}

\author[0000-0002-8465-9341]{Jacob  Burba}
\affiliation{Jodrell Bank Centre for Astrophysics, University of Manchester, Manchester, M13 9PL, United Kingdom}

\author{Steven  Carey}
\affiliation{Cavendish Astrophysics, University of Cambridge, Cambridge, UK}

\author[0000-0001-6647-3861]{Chris L. Carilli}
\affiliation{National Radio Astronomy Observatory, Socorro, NM 87801, USA}

\author[0000-0003-3197-2294]{David R. DeBoer}
\affiliation{Radio Astronomy Lab, University of California, Berkeley, CA}

\author{Eloy  de~Lera~Acedo}
\affiliation{Cavendish Astrophysics, University of Cambridge, Cambridge, UK}

\author{Matt  Dexter}
\affiliation{Radio Astronomy Lab, University of California, Berkeley, CA}

\author{Nico  Eksteen}
\affiliation{South African Radio Astronomy Observatory, Black River Park, 2 Fir Street, Observatory, Cape Town, 7925, South Africa}

\author{John  Ely}
\affiliation{Cavendish Astrophysics, University of Cambridge, Cambridge, UK}

\author[0000-0002-0086-7363]{Aaron  Ewall-Wice}
\affiliation{Department of Astronomy, University of California, Berkeley, CA}
\affiliation{Department of Physics, University of California, Berkeley, CA}

\author{Nicolas  Fagnoni}
\affiliation{Cavendish Astrophysics, University of Cambridge, Cambridge, UK}

\author[0000-0002-0658-1243]{Steven R. Furlanetto}
\affiliation{Department of Physics and Astronomy, University of California, Los Angeles, CA}

\author{Kingsley  Gale-Sides}
\affiliation{Cavendish Astrophysics, University of Cambridge, Cambridge, UK}

\author{Brian  Glendenning}
\affiliation{National Radio Astronomy Observatory, Socorro, NM}

\author[0000-0002-0829-167X]{Deepthi  Gorthi}
\affiliation{Department of Astronomy, University of California, Berkeley, CA}

\author[0000-0002-4085-2094]{Bradley  Greig}
\affiliation{School of Physics, University of Melbourne, Parkville, VIC 3010, Australia}

\author{Jasper  Grobbelaar}
\affiliation{South African Radio Astronomy Observatory, Black River Park, 2 Fir Street, Observatory, Cape Town, 7925, South Africa}

\author{Ziyaad  Halday}
\affiliation{South African Radio Astronomy Observatory, Black River Park, 2 Fir Street, Observatory, Cape Town, 7925, South Africa}

\author[0000-0001-7532-645X]{Bryna J. Hazelton}
\affiliation{Department of Physics, University of Washington, Seattle, WA}
\affiliation{eScience Institute, University of Washington, Seattle, WA}

\author{Jack  Hickish}
\affiliation{Radio Astronomy Lab, University of California, Berkeley, CA}

\author[0000-0002-0917-2269]{Daniel C. Jacobs}
\affiliation{School of Earth and Space Exploration, Arizona State University, Tempe, AZ}

\author[0000-0002-1876-272X]{Joshua  Kerrigan}
\affiliation{Department of Physics, Brown University, Providence, RI}

\author[0000-0003-0953-313X]{Piyanat  Kittiwisit}
\affiliation{Department of Physics and Astronomy,  University of Western Cape, Cape Town, 7535, South Africa}
\affiliation{South African Radio Astronomy Observatory, Black River Park, 2 Fir Street, Observatory, Cape Town, 7925, South Africa}

\author[0000-0002-2950-2974]{Matthew  Kolopanis}
\affiliation{School of Earth and Space Exploration, Arizona State University, Tempe, AZ}

\author{Adam  Lanman}
\affiliation{Department of Physics, Brown University, Providence, RI}

\author[0000-0002-4693-0102]{Paul  La~Plante}
\affiliation{Department of Astronomy, University of California, Berkeley, CA}
\affiliation{Department of Physics and Astronomy, University of Pennsylvania, Philadelphia, PA}
\affiliation{Department of Computer Science, University of Nevada, Las Vegas, NV}

\author[0000-0001-6876-0928]{Adrian  Liu}
\affiliation{Department of Astronomy, University of California, Berkeley, CA}
\affiliation{Department of Physics and McGill Space Institute, McGill University, 3600 University Street, Montreal, QC H3A 2T8, Canada}

\author{David H.~E. MacMahon}
\affiliation{Radio Astronomy Lab, University of California, Berkeley, CA}

\author{Lourence  Malan}
\affiliation{South African Radio Astronomy Observatory, Black River Park, 2 Fir Street, Observatory, Cape Town, 7925, South Africa}

\author{Cresshim  Malgas}
\affiliation{South African Radio Astronomy Observatory, Black River Park, 2 Fir Street, Observatory, Cape Town, 7925, South Africa}

\author{Keith  Malgas}
\affiliation{South African Radio Astronomy Observatory, Black River Park, 2 Fir Street, Observatory, Cape Town, 7925, South Africa}

\author{Bradley  Marero}
\affiliation{South African Radio Astronomy Observatory, Black River Park, 2 Fir Street, Observatory, Cape Town, 7925, South Africa}

\author{Zachary E. Martinot}
\affiliation{Department of Physics and Astronomy, University of Pennsylvania, Philadelphia, PA}

\author[0000-0003-3374-1772]{Andrei  Mesinger}
\affiliation{Scuola Normale Superiore, 56126 Pisa, PI, Italy}

\author{Mathakane  Molewa}
\affiliation{South African Radio Astronomy Observatory, Black River Park, 2 Fir Street, Observatory, Cape Town, 7925, South Africa}

\author[0000-0001-7694-4030]{Miguel F. Morales}
\affiliation{Department of Physics, University of Washington, Seattle, WA}

\author{Tshegofalang  Mosiane}
\affiliation{South African Radio Astronomy Observatory, Black River Park, 2 Fir Street, Observatory, Cape Town, 7925, South Africa}

\author[0000-0003-3059-3823]{Steven G. Murray}
\affiliation{Scuola Normale Superiore, 56126 Pisa, PI, Italy}
\affiliation{School of Earth and Space Exploration, Arizona State University, Tempe, AZ}

\author{Hans  Nuwegeld}
\affiliation{South African Radio Astronomy Observatory, Black River Park, 2 Fir Street, Observatory, Cape Town, 7925, South Africa}

\author[0000-0002-5400-8097]{Aaron R. Parsons}
\affiliation{Department of Astronomy, University of California, Berkeley, CA}

\author{Nima  Razavi-Ghods}
\affiliation{Cavendish Astrophysics, University of Cambridge, Cambridge, UK}

\author{James  Robnett}
\affiliation{National Radio Astronomy Observatory, Socorro, NM 87801, USA}

\author[0000-0002-2871-0413]{Peter  Sims}
\affiliation{School of Earth and Space Exploration, Arizona State University, Tempe, AZ}

\author{Craig  Smith}
\affiliation{South African Radio Astronomy Observatory, Black River Park, 2 Fir Street, Observatory, Cape Town, 7925, South Africa}

\author{Hilton  Swarts}
\affiliation{South African Radio Astronomy Observatory, Black River Park, 2 Fir Street, Observatory, Cape Town, 7925, South Africa}

\author[0000-0003-1602-7868]{Nithyanandan  Thyagarajan}
\affiliation{Commonwealth Scientific and Industrial Research Organisation (CSIRO), Space \& Astronomy, P. O. Box 1130, Bentley, WA 6102, Australia}
\affiliation{National Radio Astronomy Observatory, Socorro, NM 87801, USA}

\author{Pieter  van~Wyngaarden}
\affiliation{South African Radio Astronomy Observatory, Black River Park, 2 Fir Street, Observatory, Cape Town, 7925, South Africa}

\author{Haoxuan  Zheng}
\affiliation{Department of Physics, Massachusetts Institute of Technology, Cambridge, MA}

\date{October 2025}

\begin{abstract}
Measuring one-point statistics in redshifted 21~cm intensity maps offers an opportunity to explore non-Gaussian features of the early universe. We assess the impact of instrumental effects on measurements made with the Hydrogen Epoch of Reionization Array (HERA) by forward modeling observational and simulation data. Using HERA Phase~I observations over 94 nights, we examine the second ($m_2$, variance) and third ($m_3$) moments of images. We employ the \texttt{DAYENU}-filtering method for foreground removal and reduce simulated foreground residuals to 10\% of the 21 cm signal residuals. In noiseless cosmological simulations, the amplitudes of one-point statistics measurements are significantly reduced by the instrument response and further reduced by wedge-filtering.  Analyses with wedge-filtered observational data, along with expected noise simulations, show that systematics alter the probability distribution of the map pixels. Likelihood analysis based on the observational data shows $m_2$ measurements disfavor the cold reionization model characterized by inefficient X-ray heating, in line with other power spectra measurements. Small signals in $m_3$ due to the instrument response of the Phase~I observation and wedge-filtering make it challenging to use these non-Gaussian statistics to explore model parameters. Forecasts with the full HERA array predict high signal-to-noise ratios for $m_2$, $m_3$, and $S_3$ assuming no foregrounds, but wedge-filtering drastically reduces these ratios. This work demonstrates conclusively that a comprehensive understanding of instrumental effects on $m_2$ and $m_3$ is essential for their use as a cosmological probe, given their dependence on the underlying model.
\end{abstract}

\section{Introduction}
The epoch of reionization (EoR) marks a pivotal phase in cosmic evolution, transitioning the intergalactic medium (IGM) from mostly neutral to ionized. Studying this transition offers key insights into the formation of the first luminous sources. A primary observable from the EoR is the redshifted 21~cm line emission, originating from hyperfine splitting in ground-state neutral hydrogen. Radio interferometers, designed to detect these faint signals, are crucial for characterizing the EoR.

In recent years, various radio interferometers have undertaken extensive efforts to probe the power spectrum of 21~cm fluctuations during the EoR, leading to several upper limits. These include the Giant Metre Wave Radio Telescope \citep[GMRT;][]{Paciga2013}, the Murchison Widefield Array \citep[MWA;][]{Tingay2013, Dillon2014, Beardsley2016, Barry2019, Trott2020}, the Donald C. Backer Precision Array for Probing the Epoch of Reionization \citep[PAPER;][]{Parsons2010, Cheng2018, Kolopanis2019}, the Low Frequency Array \citep[LOFAR;][]{vanHaarlem2013, Patil2017, Mertens2020, Ghara2025}, and the Hydrogen Epoch of Reionization Array \citep[HERA;][]{Dillon2016, DeBoer2017, Berkhout2024}.

More recently, \citet[][hereafter \Ha{}]{h1c_idr2_limits} and \citet[][hereafter \Hthree{}]{h1c_idr3_limits} reported upper limits on the 21~cm power spectrum. \citet[][hereafter \Hb{}]{h1c_idr2_theory} and \Hthree{} explored the astrophysical parameter space, using these limits to constrain IGM X-ray heating and disfavor inefficient heating by $z \sim 8$.

The EoR 21~cm signal is expected to be highly non-Gaussian, especially late in reionization when ionized bubbles dominate \citep{Watkinson2014, Watkinson2015, Kittiwisit2017, Kittiwisit2022}. A straightforward way to examine non-Gaussianity in the image domain is to measure one-point statistics such as variance, skewness, and kurtosis. Theoretical studies of one-point statistics in redshifted 21~cm signals aim to link astrophysical models with the evolution of variance and skewness during reionization. For example, \citet{Wyithe2007} studied their dependence on ionizing source mass. \citet{Gluscevic2010} suggested that one-point statistics could distinguish reionization by large vs. small halos. \citet{Watkinson2014} examined these statistics under inside-out and outside-in scenarios, while \citet{Shimabukuro2015} and \citet{Watkinson2015} investigated their sensitivity to X-ray heating. These theoretical efforts support the idea that one-point statistical quantities complement power spectrum analysis and provide insights into the underlying astrophysical processes governing the reionization.

\citet{Wyithe2007} noted that an MWA-500 (500 16-dipole tiles) could detect skewness within 100–1000 hours, and an MWA-5000—with ten times the collecting area—could map skewness across redshifts. \citet{Watkinson2014} showed that 1000~hr SKA observations can distinguish their inside-out and outside-in models. \citet{Kubota2016} evaluated variance and skewness detectability with MWA and LOFAR, predicting high signal-to-noise ratios between $z \sim 7$–9 with 1000~hr exposure. These studies assumed ideal foreground removal.

\citet{Kittiwisit2017} forecasted the detectability of variance, skewness, and kurtosis with HERA in 100~hr observations, accounting for angular resolution via Gaussian kernels and frequency binning. They found that full HERA could measure one-point statistics with high sensitivity assuming perfect foreground mitigation.

In practice, measuring one-point statistics is difficult due to systematics, especially foregrounds, which are 4–5 orders of magnitude brighter than the 21~cm signal. Given the smooth spectral nature of foregrounds, several removal techniques have been developed. Blind methods like Principal Component Analysis (PCA) are effective in the image domain \citep[e.g.,][]{Alonso2015, Spinelli2022, Cunnington2022}, but may incur signal loss by removing shared components. Alternatively, Gaussian Process Regression (GPR), applied in visibility space \citep{Mertens2018}, has been refined to address this issue \citep[e.g.,][]{Mertens2020, Kern2020c, Soares2021}. A different strategy, ``foreground avoidance'' \citep[e.g.,][]{Datta2010, Vedantham2012, Morales2012, Parsons2012, Trott2012, Thyagarajan2013, Liu2014, Pober2014}, exploits the confinement of smooth-spectrum foregrounds to low Fourier modes, masking those modes to isolate the spectrally fluctuating 21~cm signal.

\citet{Kittiwisit2022} examined how foreground removal affects the measurement of one-point statistics using HERA mock observations. They adopted a foreground avoidance approach with varying filter widths. Results showed significant suppression of features linked to overdense and ionized regions. Aggressive filtering yielded washed-out intensity maps and larger uncertainties. Nonetheless, increasing trends in skewness and kurtosis toward the end of reionization were partially retained, especially with sufficient integration time.

Earlier studies on detectability of non-Gaussian signals assumed idealized maps with compact Gaussian beams. In practice, observed maps are convolved with the instrument's point spread function (PSF), resulting in dirty maps with correlated noise. The CLEAN algorithm \citep{Hogbom1974} deconvolves bright point sources but struggles with diffuse emission. Multi-resolution CLEAN methods \citep{Wakker1988, Cornwell2008b} attempt to address this, though they alter noise properties and may introduce artificial residuals, complicating statistical interpretation.

Advanced imaging methods, such as Image Domain Gridding \citep[IDG;][]{van_der_Tol2018}, improves computational efficiency by performing convolution in the image domain and handling direction-dependent effects more accurately than traditional gridding. However, IDG is not designed to preserve the statistical structure of dirty maps or model noise properties. Therefore, we choose instead to work directly with dirty maps using the direct optimal mapping, forward modeling the 21~cm signals and modeling the noise covariance matrix by propagating the measurement equation and the analysis procedure from true sky to measured image. This procedure fully takes into account the actual PSF of an interferometer.

In this study, we extend \citet{Kittiwisit2022} by incorporating realistic primary beams, instrument PSF, an advanced foreground removal method, and robust thermal noise modeling using optimal mapping \citep{Tegmark1997, Dillon2015}. We apply the Discrete Prolate Spheroidal Sequences (DPSS) foreground removal technique developed by \citet{Ewall-Wice2020}. For the first time, we measure one-point statistics from HERA Phase~I data after foreground removal, assessing the impact of residual systematics on non-Gaussian features. We also evaluate the detectability of a fiducial 21~cm model under future HERA Phase~II sensitivity.

Higher-order Fourier statistics, like the 21~cm bispectrum \citep[e.g.,][]{Yoshiura2015, Shimabukuro2016, Majumdar2018, Watkinson2018, Hutter2019, Majumdar2020, Mondal2021, Kamran2021, Watkinson2022, Raste2023}, also probe non-Gaussianity. Recent work using bispectrum \citep{Noble2024, Raste2023} and wavelet transforms \citep{Greig2023, Hothi2024} highlights their sensitivity to non-Gaussian IGM morphologies. For instance, \citet{Noble2024} shows how bispectra in specific $k$-triangle configurations distinguish inside-out, outside-in, and hybrid reionization models.

While the bispectrum shows great promise, it is also known to be susceptible to mode coupling, where different Fourier modes interact due to non-linear processes \citep{deputter2018, Shirasaki2021}. This complicates interpretation. In contrast, analyzing one-point statistics from tomographic images offers a more intuitive probe. The two approaches are complementary; here, we focus on the image domain.

This paper is structured as follows: Section~\ref{sec:dom} reviews the optimal mapping method. Section~\ref{sec:data} describes observational and simulation data. Section~\ref{sec:foreground_subtraction} outlines our advanced foreground removal technique and compares it to standard methods. Section~\ref{sec:noise_simulations} details our noise modeling. Section~\ref{sec:data_cubes} describes data cube construction. Section~\ref{sec:cc_mitigation} analyzes the impact of residual systematics post-foreground removal. Section~\ref{sec:model_comparison} compares one-point statistics from various 21cm models with HERA PhaseI data. Section~\ref{sec:detectability_H320} forecasts detectability in future observations. Finally, Section~\ref{sec:summary} summarizes our findings. Throughout, we adopt cosmological parameters from \citet{Planck2018}: $\Omega_\Lambda = 0.6844$, $\Omega_{\rm m} = 0.3156$, $\Omega_{\rm b} = 0.04911$, and $H_0 = 67.27,{\rm km, s^{-1}, Mpc^{-1}}$.

\section{Mapmaking with Direct Optimal Mapping}
\label{sec:dom}
In this section, we outline the mathematical formalism employed in mapmaking with the Direct Optimal Mapping (DOM) technique \citep{Tegmark1997, Dillon2014, Dillon2015, Xu2022, Xu2024}. DOM is considered optimal as it down-weights the impact of data with high noise variance and does not lose information about the parameters we wish to measure. DOM offers two notable advantages: a robust calculation of statistical noise for each pixel and the ability to perform mapping without assuming a flat sky. The algorithm compensates for the interferometric ``$w$-term'' exactly, without the need to perform a $w$-projection \citep{Cornwell2008a}, and it makes use of the actual $(u,v)$ -- spatial frequency -- coordinates without the need to perform $uv$-gridding and resampling; DOM directly maps visibilities to their original $uv$ locations without the need for $uv$-gridding \citep{Sullivan2012, Barry2019, Offringa2019}, thereby avoiding potential errors introduced by the gridding process. Nonzero $w$-terms can arise with wide-field mapping and with non-coplanar arrays. The HERA dishes deviate from coplanarity at the level of a few centimeters, and \citet{Xu2022} showed that neglecting such deviations result in map errors of about 5\%.

DOM requires a pixelized sky model, the main source of image infidelity (assuming perfect calibration). While DOM inversion is generally expensive, HERA’s static instrumental response (the $\mathbf{A}$ matrix) allows pre-computation, enabling efficient mapping similar to the “Fast” implementation of Fast Holographic Deconvolution \citep{Sullivan2012}.

The discretized interferometric (visibility) equation is:
\begin{align}
V_{ij}(\nu, t)= \sum^{N_{\rm src}}_{n=1} B_{ij}(\hat{s}_n, \nu) I_\nu(\hat{s}_n, \nu) e^{-2\pi i\nu \mathbf{b}_{ij}(t) \cdot \hat{s}_n/c} \Delta \Omega,
\label{eqn:rime}
\end{align}
where $i$ and $j$ denote an antenna pair, $B_{ij}$ is the peak-normalized primary beam (product of electric fields), $I_\nu$ is the sky intensity (Jy/sr), $\hat{s}_n$ the source direction, and  $\mathbf{b}(t)$ the baseline (antenna pair) vector that changes with the Earth rotation. The exponential term encodes the geometric delay (the fringe term).

For a given frequency channel, this can be written in matrix form as:
\begin{align}
\boldsymbol{\upsilon} = \mathbf{A}\mathbf{x} + \mathbf{n},
\end{align}
where $\mathbf{x}$ is the sky model vector ($N_{\rm pix}$ elements), $\boldsymbol{\upsilon}$ the observed visibilities, and $\mathbf{n}$ the thermal noise. The $\mathbf{A}$ matrix encodes beam and fringe terms, with shape $N_{\rm vis} \times N_{\rm pix}$.

Mapmaking inverts this measurement equation. The optimal estimator \citep{Tegmark1997, Dillon2015} is:
\begin{align}
\hat{\mathbf{x}} = \mathbf{D} \mathbf{A}^\dagger \mathbf{N}^{-1} \boldsymbol{\upsilon},
\label{eqn:dom}
\end{align}
where $\hat{\mathbf{x}}$ is the estimated map, $\mathbf{D}$ is a normalization matrix, and $\mathbf{N}^{-1}$ is the inverse noise covariance $\mathbf{N} = \langle \mathbf{n} \mathbf{n}^\dagger \rangle$. The expectation value of the estimator is $\langle \hat{\mathbf{x}} \rangle = \mathbf{P} \mathbf{x}$ with $\mathbf{P} = \mathbf{D} \mathbf{A}^\dagger \mathbf{N}^{-1} \mathbf{A}$. The choice of $\mathbf{N}^{-1}$ is considered, as the weighting matrix makes the map ``optimal" in the sense that it preserves information when estimating parameters from the images \citep{Tegmark1997}. Other criteria, such as a weighting that produces lower side lobes in the synthesized beam, could be incorporated into the weighting matrix instead. Since we seek to measure faint cosmological signals, we choose the $\mathbf{N}^{-1}$ weighting.

To estimate the noise matrix $\mathbf{N}$ in Equation~\eqref{eqn:dom}, we assume uncorrelated thermal noise per visibility, resulting in a diagonal matrix with elements:
\begin{align}
\sigma_{n, ij}^2 &= \frac{V_{ii} V_{jj}}{\Delta \nu \Delta t},
\label{eqn:thermal_noise}
\end{align}
where $V_{ii}$ is the autocorrelation for baseline $i$, and $\Delta \nu$, $\Delta t$ are the frequency and time resolution.

The choice of normalization matrix $\mathbf{D}$ depends on the mapmaking goal. A least-squares estimator uses $\mathbf{D} = [\mathbf{A}^\dagger \mathbf{N}^{-1} \mathbf{A}]^{-1}$ to deconvolve instrumental effects and recover the true sky. However, this matrix is often ill-conditioned, and direct inversion can amplify noise. A pseudo-inverse may help but requires careful regularization, which we do not pursue here.

We opt for the normalization scheme adopted by \citet{Xu2022}. They defined a diagonal matrix $\mathbf{D}$ with $D_{pp} = \big(\sum_q|A^\dagger_{pq}|N^{-1}_{qq}\big)^{-1} = \big(\sum_q B_{pq} N_{qq}^{-1} \big)^{-1}$ where $p$ and $q$ indicate pixel and baseline-time indices, respectively. By assuming that the noise variance of visibility does not change significantly over time, which is approximately true for slowly varying autocorrelations, the matrix can be simplified to $D_{pp} \approx \bar{\sigma}_n^2/\big(\sum_t B_{pt}\big)$ where $\bar{\sigma}_n^2 = \big(\sum_b 1/\sigma_b^2\big)^{-1}$ with baseline $b$ and $B_{pt}$ is the primary beam at pixel $p$ and time $t$. This normalization is designed to preserve the flux density of a source at zenith. 

The diagonal elements of the matrix $\mathbf{P} = \mathbf{D} \mathbf{A}^\dagger \mathbf{N}^{-1} \mathbf{A}$ become:
\begin{align}
    P_{pp} &= \sum_q D_{pp}(A^\dagger)_{pq}N_{qq}^{-1}A_{qp} \\
    &= D_{pp} \sum_q \frac{B_{pq}^2}{\sigma_q^2} \\
    &\approx \frac{\bar{\sigma}_n^2}{\sum_t B_{pt}} \frac{\sum_t B_{pt}^2}{\bar{\sigma}_n^2} = \frac{\sum_t B_{pt}^2}{\sum_t B_{pt}}.
    \label{eqn:P_matrix}
\end{align}
For example, in a single integration, the diagonal elements of the P-matrix can be approximately equal to the primary beam, meaning that the P-matrix still accounts for the primary beam and its attenuation effect. We choose not to renormalize the map to remove the weighting by the primary beam (the ``primary beam correction" of radio astronomy), because such a correction severely impacts the noise statistics in the map. For instance, introducing $D_{pp} = \big(\sum_q B_{pq}^2 N_{qq}^{-1} \big)^{-1}$ can result in $P_{pp} = 1$ when $N_{pp}$ does not change much over time, but this will diverge the noise variance at the pixel location far away from the beam center. See \citet{Xu2022} for more details.

\begin{figure}[t!]
\centering
\includegraphics[scale=0.46]{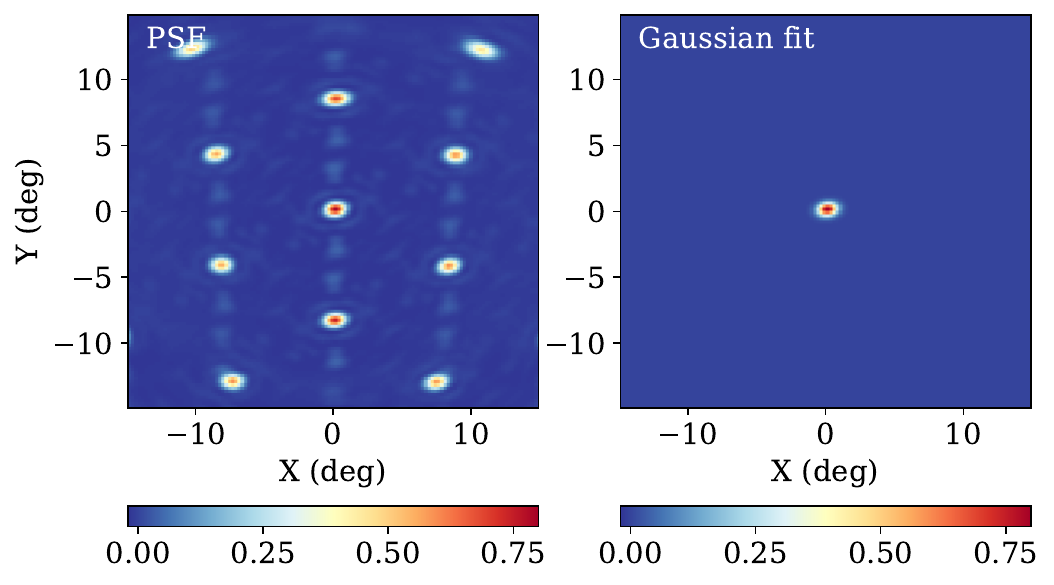}
\caption{PSF of HERA Phase~I at 161.5MHz (left) and a Gaussian fit (right). We forward-model simulated maps using the PSF to capture pixel correlations. This PSF is integrated over 1~hr. FWHM of the Gaussian fit: 1.11$^\circ$ (major) and 0.75$^\circ$ (minor).}
\label{fig:hera_psf}
\end{figure}

Figure~\ref{fig:hera_psf} shows the PSF at zenith (a column of $\mathbf{P}$ at zenith). The realistic PSF leads to inter-pixel correlations unlike Gaussian approximations commonly used. To account for this, we forward-model simulations—including 21~cm signal and noise—through the PSF to evaluate one-point statistics consistently with data.

Maps from Equation~\eqref{eqn:dom}, in Jy/beam units, are converted to mK using:
\begin{align}
\kappa(\nu) = \frac{c^2}{2k_B\nu^2 \Omega_{\rm syn}} \times 10^{-23} \quad \rm{\bigg[\frac{mK}{Jy}\bigg]},
\label{eqn:RJ_law}
\end{align}
where $c$ is the speed of light, $k_B$ is Boltzmann's constant, and $\Omega_{\rm syn} = \Omega_{\rm ant}/N_{\rm ant}$ is the synthesized beam solid angle.

HERA provides $EE$ and $NN$ polarized visibilities. We generate maps for each and form pseudo-Stokes I maps as $\hat{\mathbf{x}}_I = (\hat{\mathbf{x}}_{EE} + \hat{\mathbf{x}}_{NN})/2$. While this pseudo-Stokes I is a practical approximation to true Stokes I, it is subject to leakage from Stokes Q due to beam asymmetries. However, studies have shown that such leakage remains confined within the foreground wedge in Phase I sensitivity data \citep{Kohn2019}, and leakage-induced contamination in pseudo-Stokes I is likely subdominant to thermal noise, especially near the beam center. Therefore, for this work, we adopt pseudo-Stokes I as a valid approximation. For simulated mock observations with no systematics, we simulate $EE$ polarization only, assuming $\hat{\mathbf{x}}_I \approx \hat{\mathbf{x}}_{EE}$.

\section{Data for Phase~I Observation}
\label{sec:data}
\begin{figure*}[t!]
\centering
\includegraphics[scale=0.7]{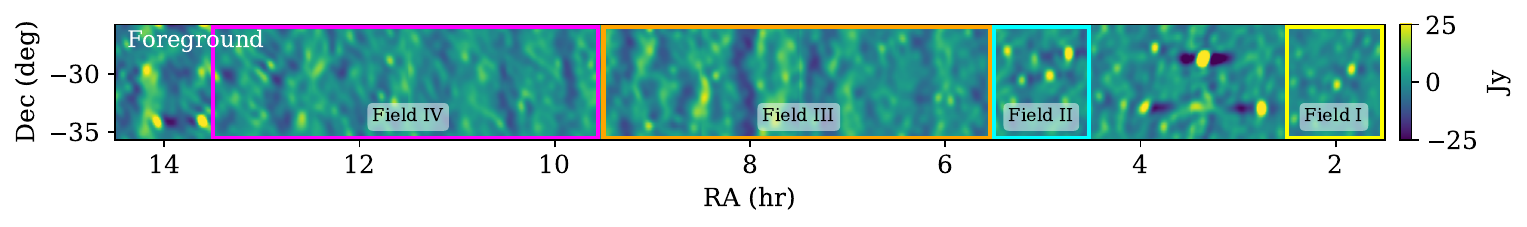}
\caption{HERA stripe, displaying foreground emission, ranging from 1.5 to 14.5 hours centered at -30.7$^\circ$ in declination. The map is created from observational data by connecting sky patches which are made at every hour of LST with the size of 15$\times$10 square degrees. Emissions at around 3~hr are grating lobes of Fornax~A, which is one of the brightest foreground radio sources.}
\label{fig:hera_stripe_foreground}
\end{figure*}

\subsection{Observational Data}
\label{sec:obs_data}
HERA is a drift-scanning interferometer in
South Africa composed of 350 14-meter dishes equipped with broad-band feeds \citep{DeBoer2017}. In this paper we make use of data collected during HERA’s first season
of operation when it was still under construction. For Phase~I the feeds consisted of crossed dipole antennas that provided sensitivity to linear polarization from 100 to 200~MHz.

Specifically, we utilize the calibrated data for Phase~I HERA observations between October 2017 and March 2018 as in \Hthree{}, and we summarize the analysis of that work here. Based on nightly inspections of the data, 94 nights were chosen to be good quality data. Each night of data were calibrated through the HERA calibration pipeline which uses redundant-baseline calibration \citep[e.g.,][]{Dillon2020} followed by absolute calibration to solve for degenerate parameters remaining after the redundant calibration \citep[e.g.,][]{Zheng2014, Dillon2018, Li2018, Kern2020a}. For LST grids with fixed bins of 21.4~s, all night observations are coherently averaged, which increases the signal-to-radios (S/N) significantly.

Before calibration antennas that are clearly not functioning are removed, and this also has the effect of removing the worst periods of interference. During calibration, solutions that fail lead to flagging of data, followed by the calculation of new solutions. Remaining artificial radio frequency interference (RFI) is then removed. However, small contributions from these regions may still exist and could introduce bias; these effects are investigated by forward modeling observational data as described in Section~\ref{sec:model_comparison}.

Two major systematics of the Phase~I observation are a cable reflection and over-the-air crosstalk between antennas, which were mitigated by \citet{Kern2019, Kern2020b}, \Ha{}, \Hthree{}. The systematics correspond to high delay modes, which can appear in foreground-removed maps. Therefore, we investigate the effect of the systematics in our mapping before and after the mitigation in Section~\ref{sec:cc_mitigation}.

The final visibility product of the pipeline for power spectrum estimation is coherently time-averaged over 214~s, but we use data without this time averaging since our mapping scheme performs an appropriate coherent averaging over time. See \Ha{} and \Hthree{} for more details about data reduction.

Out of 1024 frequency channels over 100--200~MHz, frequency ranges of 117.19--133.11~MHz and 152.25--167.97~MHz, corresponding to Band~1 and Band~2 of \Hthree{}, are considered. These bands were chosen because they were relatively RFI-free. They target the redshift of $\sim$10.4 for the low band and $\sim$7.9 for the high band.

The observation was conducted with about 40 antennas, with the actual number varying at different LSTs based on the availability of antennas at each LST. Specifically, antennas that exhibit more than 50\% good quality data over 94 nights at each LST (see ``Unflagged antennas'' in Figure~4 of \Hthree{}) were included. Timestamps with less than 10 samples are removed due to potential large error contributions.

We exclude baselines with projected east–west distances less than 14~m because the crosstalk mitigation is not applied to the baselines. The projected east-west baseline distance was chosen based on \citet{Kern2019}, where fringe-rate (FR) notch filtering was used to mitigate crosstalk by removing signals near zero FR. Short east-west baselines, which contain 21~cm signals near zero FR, suffer significant signal loss and thus are excluded.

Instead of mapping the full sky at once, we construct a series of small patches of sky, each measuring 15$\times$10 square degrees. These patches are centered at each hour of LST, ranging from 2 to 14~hr. Each patch consists of 66$\times$44 pixels and is integrated over 1.6~hr (i.e., 21.4~s$\times$268 timestamps).

To ensure accuracy, the $\mathbf{A}$ matrix is pixelized to satisfy the Nyquist criterion—sampling at least twice the PSF scale—to capture map variations without aliasing or information loss. Its sky extent is set by the primary beam, which suppresses signal far from the pointing center. Although DOM handles the $w$-term exactly and can in principle map the full sky, we restrict each mapping to a 15$\times$10 deg$^2$ region for computational efficiency. Residual effects from sources beyond this domain are expected to be negligible due to beam attenuation, though minor biases may persist.

Figure~\ref{fig:hera_stripe_foreground} shows a stitched map of sky patches. Due to the uniformity of the small patch maps, enabled by our chosen normalization scheme, the wedge-filtered signals are smoothly connected at the edges (see Figure~\ref{fig:hera_stripe_wedge_filtered}). We tested various boundary interpolation methods (none, linear, cubic spline) and found no significant impact on map properties or one-point statistics. For simplicity, we use no interpolation.

We split the stitched map into four different regions for our one-point statistics analysis: Field~I (1.5--2.5 hr), Field~II (4.5--5.5 hr), Field~III (5.5--9.5 hr), and Field~IV (9.5--13.5 hr). Each field roughly corresponds to Field~B, C, D, and E which are defined in \Hthree{}. Field~I and II are centered at 2 and 5 hr respectively, across 15 degrees, which are known to be foreground quiet radio sky. There are bright foregrounds in the 3-hr region which corresponds to grating lobes of Fornax~A, which is a bright radio galaxy located at (RA, Dec) = (3.4$^h$, -37.2$^\circ$). Thus, we do not include the region. Field~III includes the anti-galactic center, centered at $\sim$8~hr, and its sidelobes. Field~IV is the region between the anti-galactic center and galactic center.

\begin{table*}[t!]
\centering
\caption{Parameter values for different EoR models from \citet{Greig2022}.}
\label{tab:eormodels}
\resizebox{\linewidth}{!}{
\begin{tabular}{l c c c c c c c c c}
    \hline
    \textbf{Model} & $\log_{10}(f_{*, 10})$ & $\alpha_*$ & $\log_{10}(f_{\rm esc,10})$ & $\alpha_{\rm esc}$ & $t_*$ & $\log_{10} (M_{\rm turn} )$ & $\log_{10} \frac{L_{X < 2 \, \text{keV}}}{\rm SFR}$ & $E_0$ & $\alpha_{\rm X}$ \\
    & & & & & & ($M_\odot$) &  (${\rm erg\,s^{-1}\,M_\odot^{-1}\,yr}$) & (keV) & \\
    \hline
    Fiducial & -1.30 & 0.50 & -1.00 & -0.50 & 0.50 & 8.7 & 40.50 & 0.50 & 1.00\\
    Cold reionization & -1.30 & 0.50 & -1.00 & -0.50 & 0.50 & 8.7 & 38.00 & 0.50 & 1.00 \\
    Large halos & -0.70 & 0.50 & -1.00 & -0.50 & 0.50 & 9.9 & 40.50 & 0.50 & 1.00 \\
    Extended reionization & -1.65 & 0.50& -1.00 & -0.50 & 0.50 & 8.0 & 40.50 & 0.50 & 1.00 \\
    \hline
\end{tabular}}
\end{table*}

\subsection{Simulation Data}
\subsubsection{21~cm Signal Simulations}
\label{sec:eor_simulations}
The brightness temperature of 21~cm signals is given by the contrast between the 21~cm spin temperature and the background radiation temperature (i.e., CMB temperature), which can be written as \citep[e.g.,][]{Furlanetto2006},
\begin{align}
    \delta T_b(\nu) &= \frac{T_{\rm S} - T_\gamma}{1+z} \big(1-e^{-\tau_{21}}\big) \nonumber \\
    &\approx 27 x_{\rm H_I} (1 + \delta_{\rm m}) \bigg(\frac{H}{dv_\parallel/dr + H}\bigg) \bigg(1-\frac{T_\gamma}{T_{\rm S}}\bigg) \nonumber \\
    &\;\times \bigg(\frac{1+z}{10}\frac{0.15}{\Omega_{\rm M}h^2}\bigg)^{1/2} \bigg(\frac{\Omega_{\rm b}h^2}{0.023}\bigg) \; {\rm mK},
    \label{eqn:21cm_signal}
\end{align}
where $T_{\rm S}$ is the gas spin temperature, $T_\gamma$ is the CMB temperature, $\tau_{21}$ is the optical depth at the 21~cm frequency, $x_{\rm H_I}$ is the neutral fraction of the hydrogen gas, and $\delta_{\rm m} = \rho/\Bar{\rho}-1$ is the matter density fluctuations where $\rho$ and $\Bar{\rho}$ are the matter density and its mean, respectively. $dv_\parallel/dr$ is the gradient of the peculiar velocity along the line-of-sight. This equation provides the baseline for computing the 21~cm brightness temperature.

We generate 21~cm brightness temperature cubes and associated astrophysical fields using the publicly available code \texttt{21cmFAST} \citep{Mesinger2011, Murray2020}. To study the non-Gaussian properties of the 21cm signal under varying astrophysical conditions, we simulate four reionization scenarios: a fiducial model from \citet{Park2019} and three alternatives proposed by \citet{Greig2022}—cold reionization, large halos, and extended reionization.

The fiducial model is calibrated to match the high-redshift galaxy UV luminosity function and mock 21cm observations. The cold reionization scenario suppresses X-ray heating of the IGM, resulting in stronger 21cm signal fluctuations compared to the fiducial model; however, this scenario is disfavored by recent power spectrum constraints (\Hb{}, \Hthree{}) \citep{Mesinger2014, Parsons2014}. The large halos model assumes a higher characteristic halo mass ($M_{\rm turn}$), limiting star formation to massive halos and making reionization primarily driven by bright galaxies. In contrast, the extended reionization scenario adopts a lower star formation efficiency, delaying reionization and shifting the dominant contribution to faint galaxies.

Key model parameters are summarized in Table~\ref{tab:eormodels}, including the stellar mass normalization $f_{,10}$, turnover halo mass $M_{\rm turn}$, and the soft X-ray luminosity per unit star formation rate $L_{X<2 \text{keV}}/\mathrm{SFR}$. Additional parameters—such as the stellar mass power-law slope ($\alpha_*$), star formation timescale ($t_*$), ionizing photon escape fraction parameters ($f_{\rm esc,10}$, $\alpha_{\rm esc}$), X-ray spectral index ($\alpha_{\rm X}$), and the minimum X-ray energy ($E_0$)—are held fixed across models as shown in Table~\ref{tab:eormodels}.

To reduce computational cost, we omit the subgrid recombination model of \citet{Park2019} and instead fix the mean free path parameter, $\texttt{R\_BUBBLE\_MAX}$, to 15~Mpc in all simulations. In addition, unlike previous studies, we rely on halo mass derived from a halo finder in computing ionization fields by setting \texttt{USE\_HALO\_FIELD = True}.

The models are simulated over a redshift range of 6.5--13.0 with a cube size of $200^3$ Mpc$^3$ with 128 pixels per side. The coeval cubes are then interpolated to a whole sky map using Hierarchical Equal Area isoLatitude Pixelization \citep[\texttt{HEALPix};][]{Gorski2005} at the redshift of interest, forming a lightcone data over redshifts. We interpolate the cubes into \texttt{nside} = 8192 to minimize interpolation artifacts and down-sample it to a computationally feasible \texttt{nside} = 128. The details of this process are described in Appendix~A of \citet{Kittiwisit2017}.

A statistical moment is a quantitative measure that provides information about the shape and characteristics of a distribution of data. The $p$-th moment is defined as
\begin{align}
    m_p = \frac{1}{N} \sum_{i=1}^N (x_i - \mu)^p,
    \label{eqn:moments}
\end{align}
where $N$ denotes the number of data, and $\mu$ represents the mean of the dataset.

Specific quantities we are interested in are variance and skewness. The variance measures the dispersion of the data which is closely related to a power spectrum. The skewness provides information about the asymmetry of the distribution that potentially captures non-Gaussianity. Mathematically, they are expressed as
\begin{align}
    S_2 &= m_2 \label{eqn:variance}, \\
    S_3 &= m_3 / (m_2)^{3/2} \label{eqn:skewness},
\end{align}
where $m_2$ and $m_3$ are the second and third moments, and $S_2$ and $S_3$ represent variance and skewness, respectively. Since our analysis focuses on small sky patches where the noise variance is nearly uniform across pixels, we do not incorporate additional per-pixel weighting when calculating those statistical measures.

\begin{figure}[t!]
\centering
\includegraphics[scale=0.35]{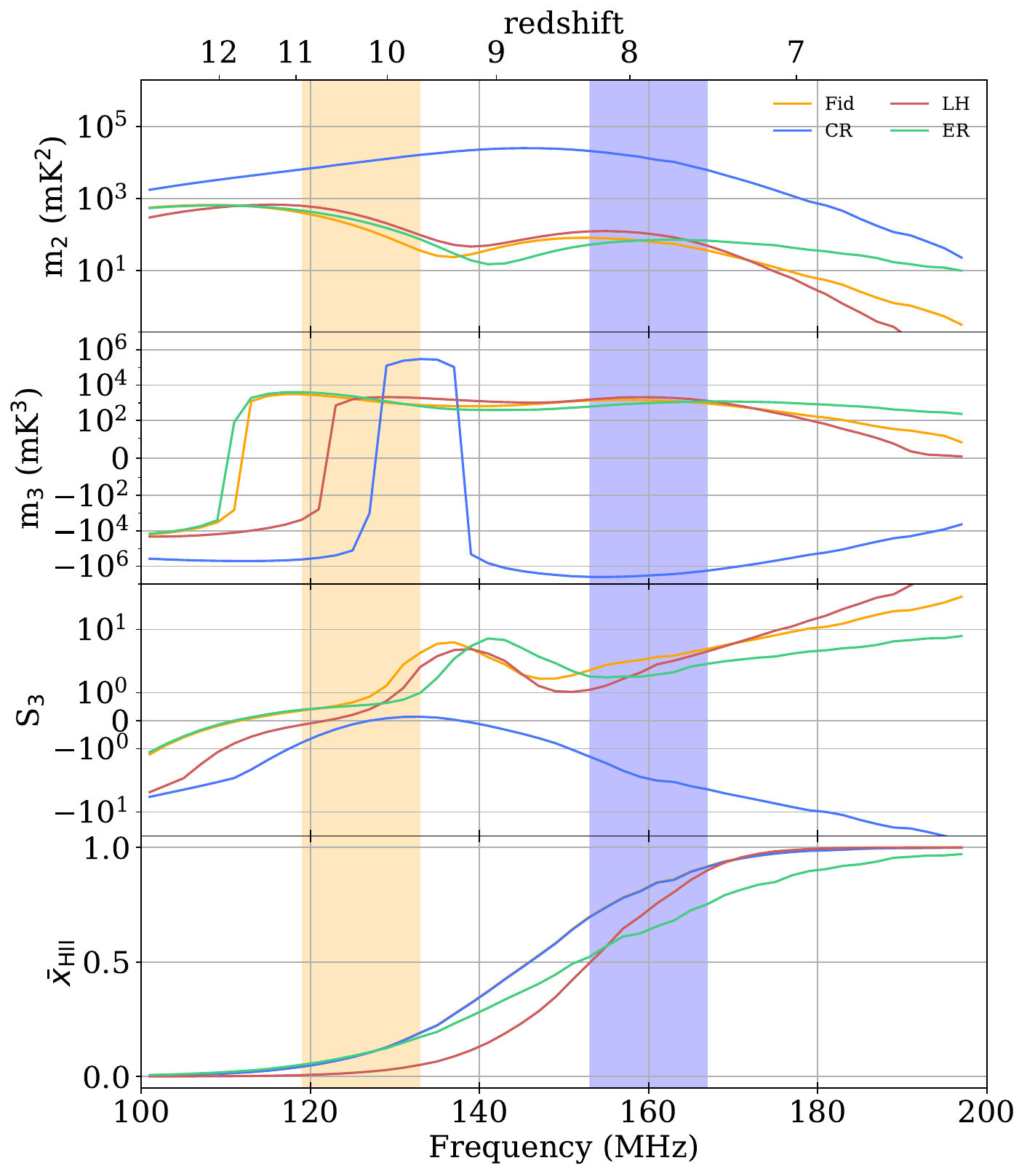}
\caption{Evolutionary history of $m_2$ (equivalent to $S_2$), $m_3$, $S_3$, and mean ionization fraction measured from raw simulations for the four distinct 21~cm models. Each line indicates the fiducial (Fid), cold reionization (CR), large halos (LH), and extended reionization (ER) models. Band~1 and Band~2 are highlighted in orange and blue shaded regions.}
\label{fig:EoR_models}
\end{figure}

\begin{figure*}[t!]
\centering
\includegraphics[scale=0.45]{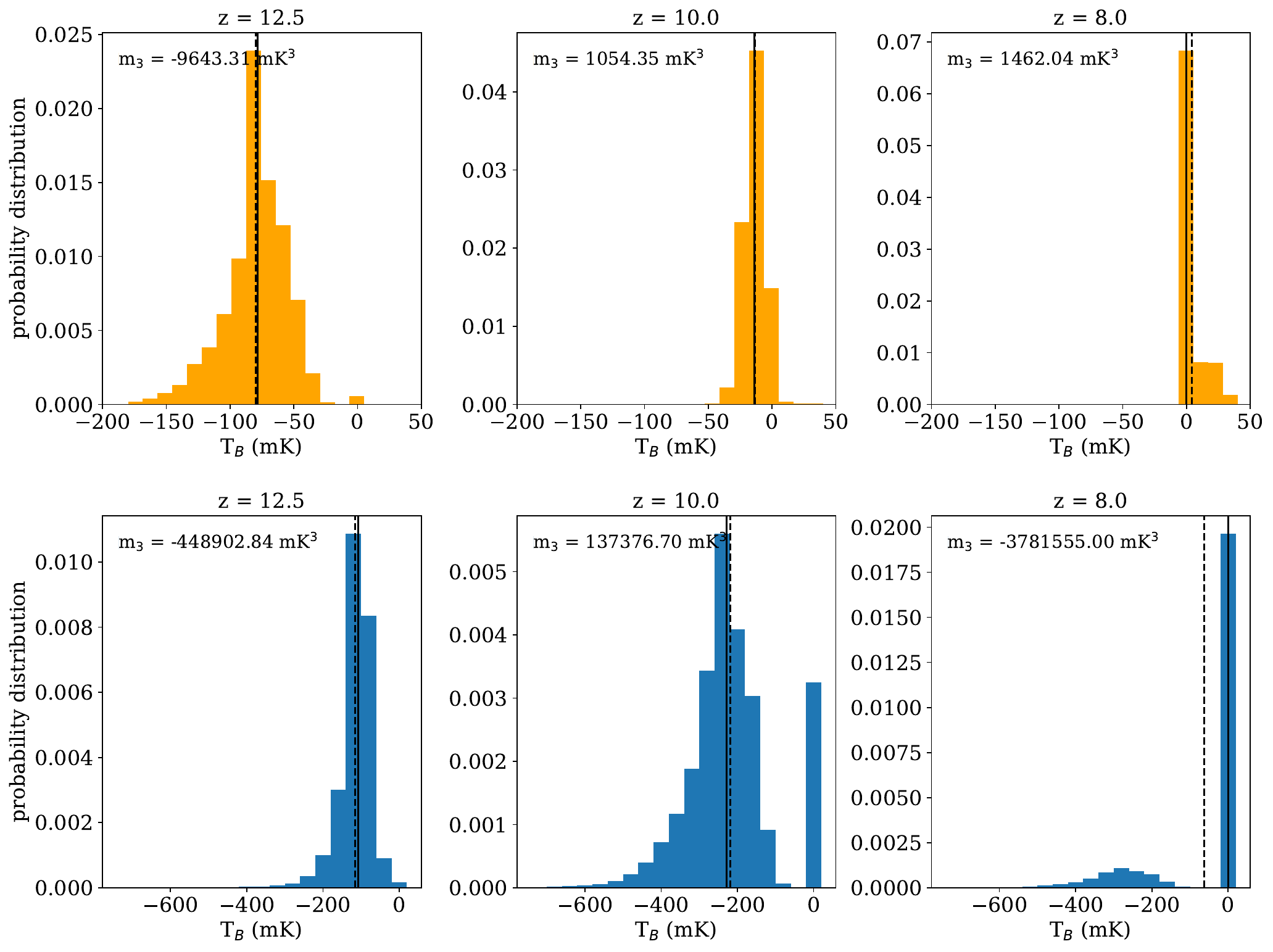}
\caption{Probability distribution of the fiducial (top row) and cold reionization (bottom row) models at specific redshifts for raw simulations (i.e., no instrumental effect). The dashed and solid vertical lines indicate the mean and median, respectively. At $z = 12.5$, both models exhibit asymmetric distributions due to heating, leading to a negative $m_3$. As reionization advances, a prominent pileup at the zero bin becomes noticeable. In the fiducial model, efficient heating by star-forming galaxies raises the temperature of the IGM, resulting in a tail towards positive values and thus a positive $m_3$ (top right panel). In the cold reionization model, the distribution at $z = 10.0$ centered around T$_B \sim -250$~mK exhibits negative tails. However, $m_3$ transitions to positive values due to the ionized IGM at the zero bin (bottom center panel). As reionization progresses further, $m_3$ becomes negative again as a large fraction of the IGM is ionized, leaving behind some negative IGM regions (bottom right panel).}
\label{fig:P_D_raw}
\end{figure*}

Figure~\ref{fig:EoR_models} shows the evolution of the second moment (first panel), third moment (second panel), skewness (third panel), and mean ionization fraction (last panel) measured from raw simulations (i.e., no instrumental effect) for each model. The cold reionization model shows significantly higher variance than the others, driven by its low spin temperature (see Equation~\eqref{eqn:21cm_signal}), making it especially relevant near current sensitivity limits. Skewness, a key indicator of non-Gaussianity, grows sharply at low redshifts as ionization from star-forming galaxies increases. In the cold reionization model, skewness trends negative due to a spin temperature below the CMB temperature. Differences in reionization history (e.g., driven by bright galaxies in the large-halos model versus faint galaxies in the extended reionization model) manifest in both variance and skewness evolution, offering potential observables to distinguish between scenarios.

The fiducial model exhibits a sign flip in $m_3$ around $z \sim 12$, which can be understood by examining the evolution of the brightness temperature distribution (top panels of Figure~\ref{fig:P_D_raw}). At $z \sim 12.5$, the distribution shows a negative tail due to early heating, resulting in $m_3 < 0$. As the IGM warms and ionizes, the distribution shifts toward a positive tail with a mean near 0$\sim$mK, producing $m_3 > 0$.

A distinct feature appears in the cold reionization model (Figure~\ref{fig:EoR_models}), where $m_3$ undergoes two sign flips, unlike the other scenarios. As shown in the bottom panels of Figure~\ref{fig:P_D_raw}, the model initially follows a similar path, with a negatively skewed distribution at $z \sim 12.5$. However, as reionization progresses with minimal heating, the zero bin height increases, driving $m_3$ positive, an effect we refer to as sign-flip instability. Eventually, once a large portion of the IGM is ionized, residual neutral regions with negative brightness temperatures cause $m_3$ to turn negative again.

\begin{table*}[t!]
    \centering
    \caption{Summary of Observational and Simulation Data for mapmaking.}
    \label{tab:summary_data}
    \resizebox{\linewidth}{!}{
    \begin{tabular}{c c c}
        \hline
        Data & Description & Field \\
        \hline
        \textbf{Observation (Phase~I)} & With Systematics Mitigation & I, II, III, IV \\
        & Without Systematics Mitigation & I, II, III, IV \\
        \textbf{Simulation (Phase~I)} & & \\
        Foreground + 21~cm (fiducial) & Foreground Subtraction Validation & II \\
        21~cm (four models) & EoR Detection Test Model & I, II, III, IV \\
        21~cm (fiducial \& cold reionization) & Sample Variance Estimation & I, II, III, IV \\
        Thermal Noise & Noise limit \& Uncertainty of Mock Observation & I, II, III, IV \\
        \textbf{Simulation (Phase~II)} & & \\
        21~cm (fiducial) & Forecasting Future Observation & 1.5--5.5 hr \\
        Thermal Noise & Sensitivity on Prediction & 1.5--5.5 hr \\
        \hline
    \end{tabular}}
\end{table*}

The simulated whole sky maps are then transformed into our mock radio observation using Equation~\eqref{eqn:rime}\footnote{We utilize a simulator provided by \url{https://github.com/vispb/vispb}.}. The mock observations are based on the antenna layout of the HERA Phase~I observation (\Hthree{}) using a dipole feed beam for the primary beam \citep{Fagnoni2020}. The simulated EoR signals described above are employed for the intensity of sources in Equation~\eqref{eqn:rime}. The visibility simulations cover a frequency range of 115--190~MHz, including Band~1 and Band~2, with a spacing of 97.7~kHz. The mock observations span a 0--16~hr LST window that covers the fields of interests as described in Section~\ref{sec:obs_data}, at a time cadence of 21.4~seconds.

In order to account for the sample variance in one-point statistics measurements, we sample the sky at 30 different locations and construct mock visibilities for the fiducial and cold reionization model.

\subsubsection{Foreground Simulations}
To assess the performance of foreground removal techniques in Section~\ref{sec:foreground_subtraction}, we also create foreground visibilities by using Equation~\eqref{eqn:rime}. We choose Field~II centered at 5~hr for the test of foreground removal. For the foreground sources, we include compact radio components and diffuse sky models. The former is drawn from the GaLactic and Extragalactic All-sky MWA \citep[GLEAM;][]{Hurley-Walker2017} survey, and the latter is from a Global Sky Model \citep[GSM;][]{Zheng2017}. 

For the point-source sky model, we incorporate both GLEAM~I \citep{Hurley-Walker2017} and GLEAM~II \citep{Hurley-Walker2019} catalogs, along with the peeled bright sources documented in Table 2 of \citet{Hurley-Walker2017} and Fornax A \citep[e.g.,][]{Bernardi2013}. The GLEAM~I survey essentially spans the entire southern hemisphere and achieves 95\% completeness at 160~mJy, excluding regions corresponding to the Large and Small Magellanic Clouds and the galactic plane stripe. The GLEAM~II catalog aims to fill the gap within the diffuse galactic plane. While the second catalog is not entirely complete, observations at LST~=~5~hr are minimally affected by this incompleteness, as the primary beam's main lobe is sufficiently distant from the galactic plane. In addition, we try to fill in the gap using a GSM described below.

\citet{Zheng2017} provides diffuse GSM maps with the bright compact sources removed from the sky model. This precaution helps minimize the risk of double-counting bright point sources when merging the point-source and diffuse sky models. The GSM offers high-resolution models in \texttt{HEALPix}, pixelized at \texttt{nside} = 1024 for each frequency. The maps are then downgraded to \texttt{nside} = 256 to ensure computational feasibility. The combination of the point source survey and GSM equips us adequately to investigate the impact of foregrounds on our foreground subtraction techniques, when the simulated foreground visibilities are added to the EoR visibility simulations. The summary of the data used in this study is described in Table~\ref{tab:summary_data}.

\section{Foreground Removal}
\label{sec:foreground_subtraction}
\label{sec:wedge_filter}
In this section, we discuss a foreground filtering method using DPSS within the context of wedge-filtering.

Wedge-filtering is a tool to avoid the foreground that occupies low Fourier modes and to extract cosmological signals located at high delay modes. Specifically, we employ the 2D power spectrum defined by the Fourier modes $k_\perp$ and $k_\parallel$, perpendicular and parallel to the line-of-sight, respectively. The foreground with smooth spectral structure tends to be located at low $k_\parallel$ in the 2D power spectrum. The size of the foreground region along the $k_\parallel$ axis in the power spectrum increases with $k_\perp$ due to the chromaticity of the interferometer, forming the foreground wedge. Because the 21~cm signals are expected to have complex spectral structure, they can exist outside the foreground wedge, called the EoR window \citep{Datta2010, Vedantham2012, Morales2012, Morales2018, Parsons2012, Trott2012, Thyagarajan2013, Liu2014, Pober2014}.

The wedge-filtering is then selecting $k_\|$ modes in the EoR window, larger than a criterion. The criterion can be chosen based on the field-of-view (FoV) of an instrument that determines a maximum $k_\parallel$ or delay mode occupied by foregrounds, $\tau_{\rm max} = b\sin{\theta}/c$ where $b$ is a separation between an antenna pair, $\theta$ is the angular extent of a dish, and $c$ is the speed of light \citep{Parsons2012, Liu2014}. Note that $k_\parallel \propto \tau$.

For HERA 21~cm cosmology measurements, we must consider the FoV to be down to the horizon of the observer's frame, and thus the wedge-filtering size is subject to the horizon limit ($\tau_{\rm horizon} = b/c$). In addition, we may need to apply an extra buffer ($\tau_{\rm buffer}$) above the horizon limit due to the chromaticity of the primary beam, especially when there is strong horizon emission convolved with the primary beam \citep{Thyagarajan2016} as well as systematics such as calibration errors \citep{Orosz2019, Kim2022, Kim2023}. For instance, $\tau_{\rm buffer} = 300$~ns is considered in \Hthree{} for their power spectrum estimation. In this study, we choose the same buffer size of 300~ns for the wedge-filtering.

A specific implementation of wedge-filtering involves constructing an inverse covariance matrix in delay (Fourier) space using a top-hat filter, as proposed by \citet{Ewall-Wice2020}. Their linear foreground filtering method is based on the Discrete Prolate Spheroidal Sequences (DPSS) basis, which has two key properties: (1) it diagonalizes the covariance matrix defined by the top-hat filter, corresponding to a sinc function (${\rm sinc}\,x = \sin{x}/x$) in frequency space, and (2) it concentrates foreground power within the filter window in Fourier space. These features make the method effective and well-suited for isolating the 21~cm signal.

We utilize the DPSS Approximate lazY filtEriNg of foregroUnds \citep[\texttt{DAYENU};][]{Ewall-Wice2020} method\footnote{The python library is publicly available in \url{https://github.com/HERA-Team/uvtools}.} to filter out foregrounds per baseline. The filtering process is governed by two essential parameters: the filtering size and a suppression factor for foregrounds within the filtering window. Due to the sub-optimal performance at the band edges, we exclude the first and last four frequency channels in our analysis.

The wedge-filtered visibilities are turned into images using Equation~\eqref{eqn:dom}. By exploring simulated foregrounds on top of fiducial EoR simulations without thermal noise, we find that a filtering size of $\tau_{\rm filter} = \tau_{\rm horizon} + 300$~ns and a suppression factor of $\epsilon = 10^{-9}$ are sufficient to reduce foreground residuals to levels compatible with EoR signal detection (i.e., foreground residuals at about 10\% of the 21 cm signal residuals), consistent with the results of \citet{Ewall-Wice2020}. Furthermore, \citet{Ewall-Wice2020} demonstrated that \texttt{DAYENU} efficiently suppresses smooth foregrounds while leaving most of the noise unaffected, ensuring that the wedge-filtering remains valid even with additional noise present. Based on this, we apply the filtering size and suppression factor to the EoR-only simulations in Sections~\ref{sec:model_comparison} and \ref{sec:detectability_H320}, assuming residual foregrounds in simulations are adequately suppressed.

Another technique for wedge-filtering involves masking low-delay modes containing foregrounds with the Fourier basis adopted by other studies \citep[e.g.,][]{Prelogovic2021, Kittiwisit2022}. For instance, to mitigate the edge effects caused by applying a tapering function in their wedge-filtering, \citet{Kittiwisit2022} employed a rolling-window method. While this tapering function helps reduce side lobes of foregrounds in Fourier space, some foreground emission can still leak into the EoR window after filtering. Since \texttt{DAYENU} does not rely on any tapering function, it avoids edge effects and outperforms the conventional wedge-masking technique. Therefore, in the following sections, we adopt the \texttt{DAYENU}-filtering method for our wedge-filtering approach.

\begin{figure}[t!]
\centering
\includegraphics[scale=0.35]{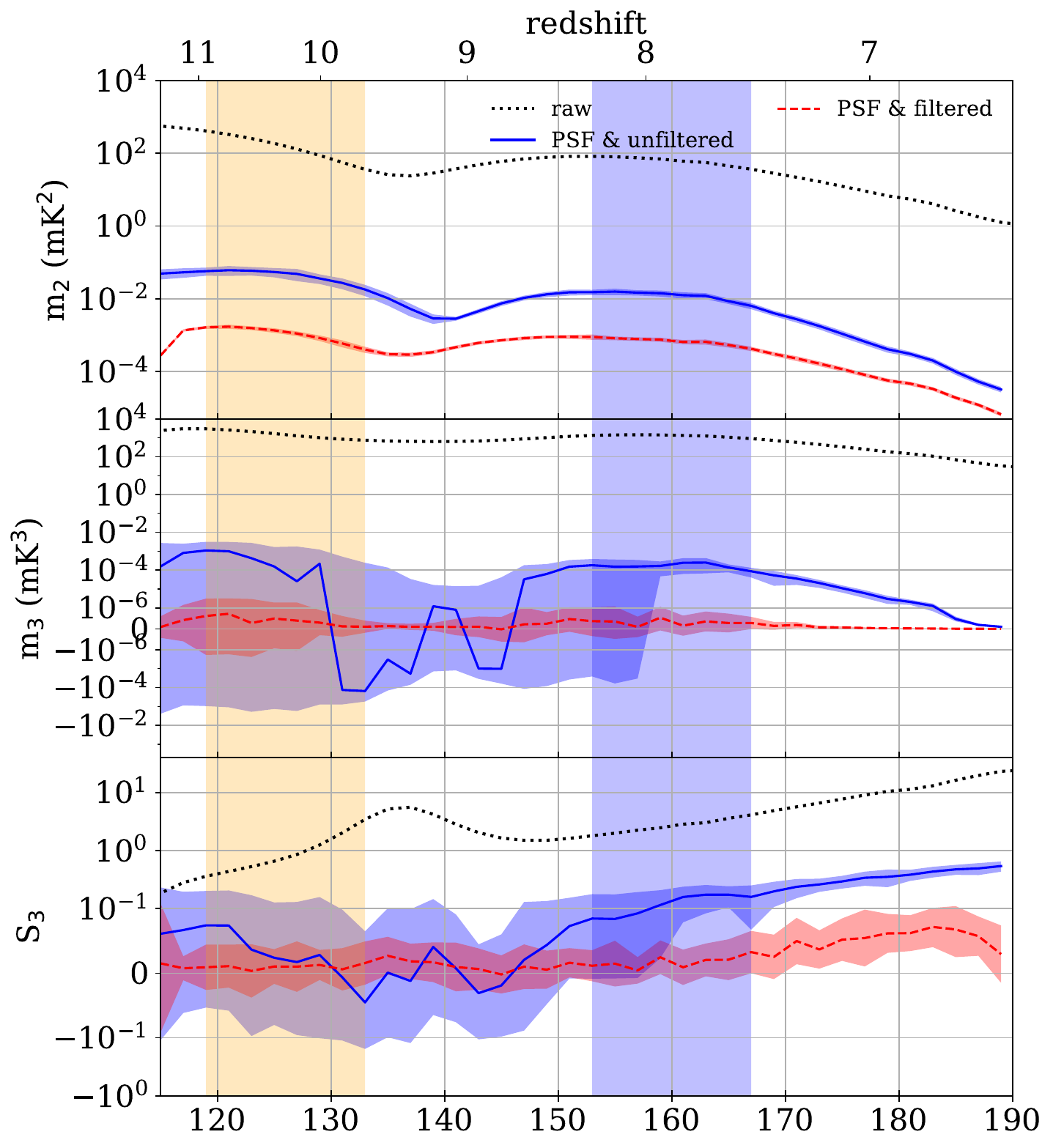}
\caption{$m_2$ (top), $m_3$ (middle), and $S_3$ (bottom) for the fiducial model as a function of frequency given Phase~I mock observations. $m_2$ (top), $m_3$ (middle), and $S_3$ (bottom) for the fiducial model are shown as functions of frequency, based on Phase~I mock observations. The dotted line represents measurements from raw simulations without instrumental effects. The blue solid line and shaded region show the mean and sample variance after applying the PSF, assuming no foregrounds. The red dashed line and shaded region show results after applying both the PSF and wedge-filtering. The PSF significantly reduces the amplitudes of $m_2$ and $m_3$, with wedge-filtering further suppressing $m_3$, making detection of non-Gaussian features more difficult. For $m_3$, a linear scale is used between $-10^{-6}$ and $10^{-6}$; outside this range, the scale is logarithmic. For $S_3$, the linear range spans $-10^{-1}$ to $10^{-1}$, with logarithmic scaling beyond.}
\label{fig:1pnt_psf_filtering_effect_fiducial}
\end{figure}

\begin{figure}[t!]
\centering
\includegraphics[scale=0.35]{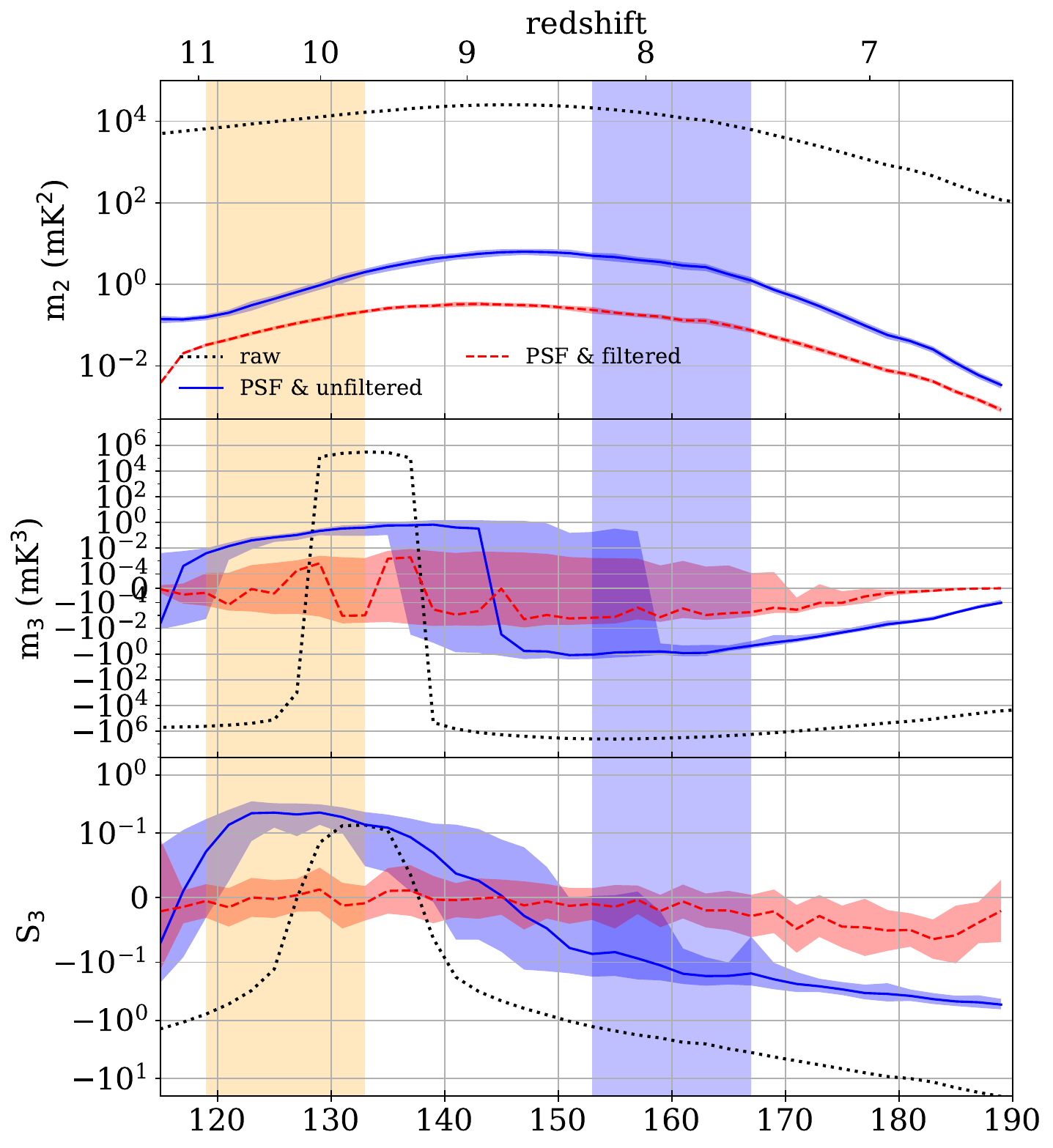}
\caption{$m_2$ (top), $m_3$ (middle), and $S_3$ (bottom) for the cold reionization model as a function of frequency given Phase~I mock observations. The format is the same as that of Figure~\ref{fig:1pnt_psf_filtering_effect_fiducial} but for the cold reionization model. Similar to the fiducial model, the variance is reduced by a factor of $10^4$ when the PSF is applied, and it is further reduced by two orders of magnitude after wedge-filtering. As with the fiducial model, detecting non-Gaussianity, whether through $m_3$ or $S_3$, becomes challenging due to the combined effects of the PSF and wedge-filtering. For $m_3$, the scale is linear between $-10^{-4}$ and $10^{-4}$, transitioning to a logarithmic scale outside this range. For $S_3$, the scale is linear between $-10^{-1}$ and $10^{-1}$, with a logarithmic scale applied beyond these limits.}
\label{fig:1pnt_psf_filtering_effect_cold_reionization}
\end{figure}

\begin{figure}[t!]
\centering
\includegraphics[scale=0.35]{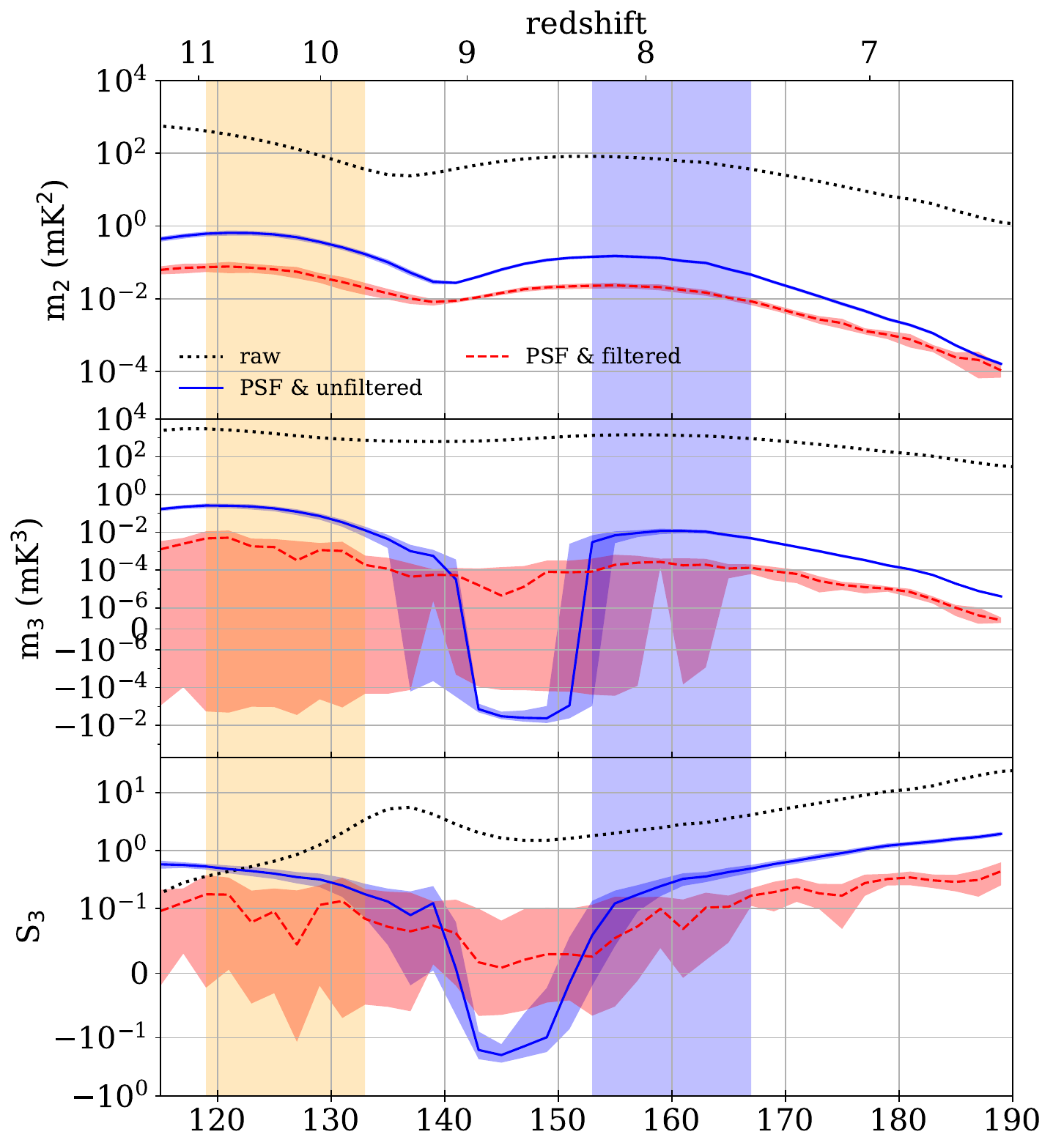}
\caption{$m_2$ (top), $m_3$ (middle), and $S_3$ (bottom) for the fiducial model as a function of frequency with HERA-320. The format is the same as that of Figure~\ref{fig:1pnt_psf_filtering_effect_fiducial} but for the HERA-320 configuration. The PSF effect reduces $m_2$ by a factor of $10^3$ assuming no foreground, with further reduction through wedge-filtering by an order of magnitude. In the blue curve, $m_3$ exhibits a distinct trend compared to Figure~\ref{fig:1pnt_psf_filtering_effect_fiducial}, with relatively smaller sample variance. The red line shows that wedge-filtering decreases $m_3$ signals and increases the sample variance, making it challenging to detect significant $m_3$ signals at $z > 7.5$. However, there may still be potential to detect deviations in $m_3$ and $S_3$ measurements from the zero baseline at low redshifts.}
\label{fig:1pnt_psf_filtering_effect_fiducial_h320}
\end{figure}

\section{instrumental effects on One-point Statistics}
\label{sec:instrument_effects_1pnt_stat}

In Section~\ref{sec:foreground_subtraction}, we explored the effect of wedge-filtering on the resultant images. In this section, we examine how the PSF and wedge-filtering affect measuring one-point statistics of our theory models. Figure~\ref{fig:1pnt_psf_filtering_effect_fiducial} shows $m_2$ (top), $m_3$ (middle), and $S_3$ (bottom) measurements for the (thermal) noiseless fiducial model. The dotted lines indicate raw EoR model which means it is free from the effect of instrument and foregrounds.

The blue lines represent cases where the instrument response or PSF is applied without accounting for foreground contamination. The solid lines and shaded regions denote the mean and sample variance measured across a $60\times10$ square degree field over 30 different sky realizations, respectively. Due to our map normalization scheme, which preserves surface brightness, the one-point statistics measurements are not conserved after PSF convolution, leading to a significant reduction in variance by a factor of $10^4$. Although $m_3$ is affected by large sample variance, there remains a possibility of detecting non-zero $m_3$ signals at higher frequency bands. The reduced sample variance beyond 160 MHz may result from the increased reionization of neutral hydrogen, causing different sky fields to appear more similar. Additionally, $S_3$ (bottom panel) suggests that meaningful detection at $z > 8$ is challenging, but there is an increasing trend in $S_3$ toward lower redshifts, despite its magnitude being reduced by roughly an order of magnitude compared to $S_3$ derived from the raw simulation.

When foregrounds are accounted for and wedge-filtering is applied to mitigate them, the amplitude of the one-point statistics measurements is further reduced. The dashed red line in Figure~\ref{fig:1pnt_psf_filtering_effect_fiducial} shows that $m_2$ decreases by an order of magnitude, and the signal in $m_3$ approaches zero, indicating that most of the non-Gaussianity information is washed out due to the wedge-filtering. The $S_3$ plot suggests that detecting non-zero skewness in Band~1 and Band~2 is challenging, although there is a slight increasing trend toward lower redshifts.

Figure~\ref{fig:1pnt_psf_filtering_effect_cold_reionization} presents the results for the cold reionization model. Similar to the fiducial model, $m_2$ decreases by a factor of $10^4$ after the instrument response is applied and is further reduced by an additional order of magnitude after the wedge-filtering is applied.

In Section~\ref{sec:eor_simulations}, we discussed the sign flips in $m_3$ observed in the raw simulations of the cold reionization model, which are revisited by the black dotted line in Figure~\ref{fig:1pnt_psf_filtering_effect_cold_reionization}. The sign flips were previously interpreted based on the idea that the skewed shape of the brightness temperature distribution, combined with the pileup in the zero bin due to reionization, determines the sign of $m_3$, as illustrated in Figure~\ref{fig:P_D_raw}. When considering the PSF effect with no foreground (blue solid line), sign flips still occur but at different locations compared to the black dotted line, possibly due to additional effects like grating and/or side lobes. The blue shaded region indicates that the uncertainty in the location of these sign flips spans 140–-160 MHz. However, once wedge-filtering is applied, the $m_3$ and $S_3$ signals are mostly removed.

In addition to the Phase I observation, we explore the predicted one-point statistics for the HERA-320 array configuration. The PSF of HERA-320 is narrower compared to that of the Phase I observation, allowing us to probe smaller scales. Figure~\ref{fig:1pnt_psf_filtering_effect_fiducial_h320} shows the results for the fiducial model with 320 antennas. The mean and sample variance, represented by lines and shaded regions respectively, are estimated across a $60 \times 10$ square degree field from 30 random independent sky locations. Due to the ability to capture additional signals from smaller scales, the variance after applying the PSF (blue line) is larger than in Figure~\ref{fig:1pnt_psf_filtering_effect_fiducial}, though still reduced by a factor of $10^3$ compared to the raw simulations. The wedge-filtering continues to reduce the variance by an order of magnitude, as indicated by the red curve.

Unlike the Phase I observation, which utilized around 40 antennas, HERA-320 predicts that non-zero $m_3$ signals (assuming no foregrounds, as shown by the blue line) are detectable across redshifts, with sign flips being relatively well-identified due to the smaller sample variance compared to Figure~\ref{fig:1pnt_psf_filtering_effect_fiducial}. However, when wedge-filtering is applied, detecting non-zero $m_3$ at $z > 7.5$ becomes challenging because of the large sample variance. Nonetheless, there remains some potential for detecting non-zero $m_3$ and $S_3$ at lower redshifts in the red curves. To fully assess detectability for future observations, it is necessary to account for thermal noise in addition to sample variance. In Section~\ref{sec:detectability_H320}, we explore the detectability with HERA-320 over redshifts, considering 300 night observation time.

\section{Noise Simulations}
\label{sec:noise_simulations}
If the observational data is free from systematics and dominated by thermal noise after foreground removal, noise simulations provide theoretical expectation of measured one-point statistics (Sections~\ref{sec:cc_mitigation} and \ref{sec:model_comparison}). Additionally, noise simulations are used to indicate the sensitivity of future observations, allowing us to forecast one-point statistics for the full core array of HERA (Section~\ref{sec:detectability_H320}).

\begin{figure*}[t!]
\centering
\includegraphics[scale=0.4]{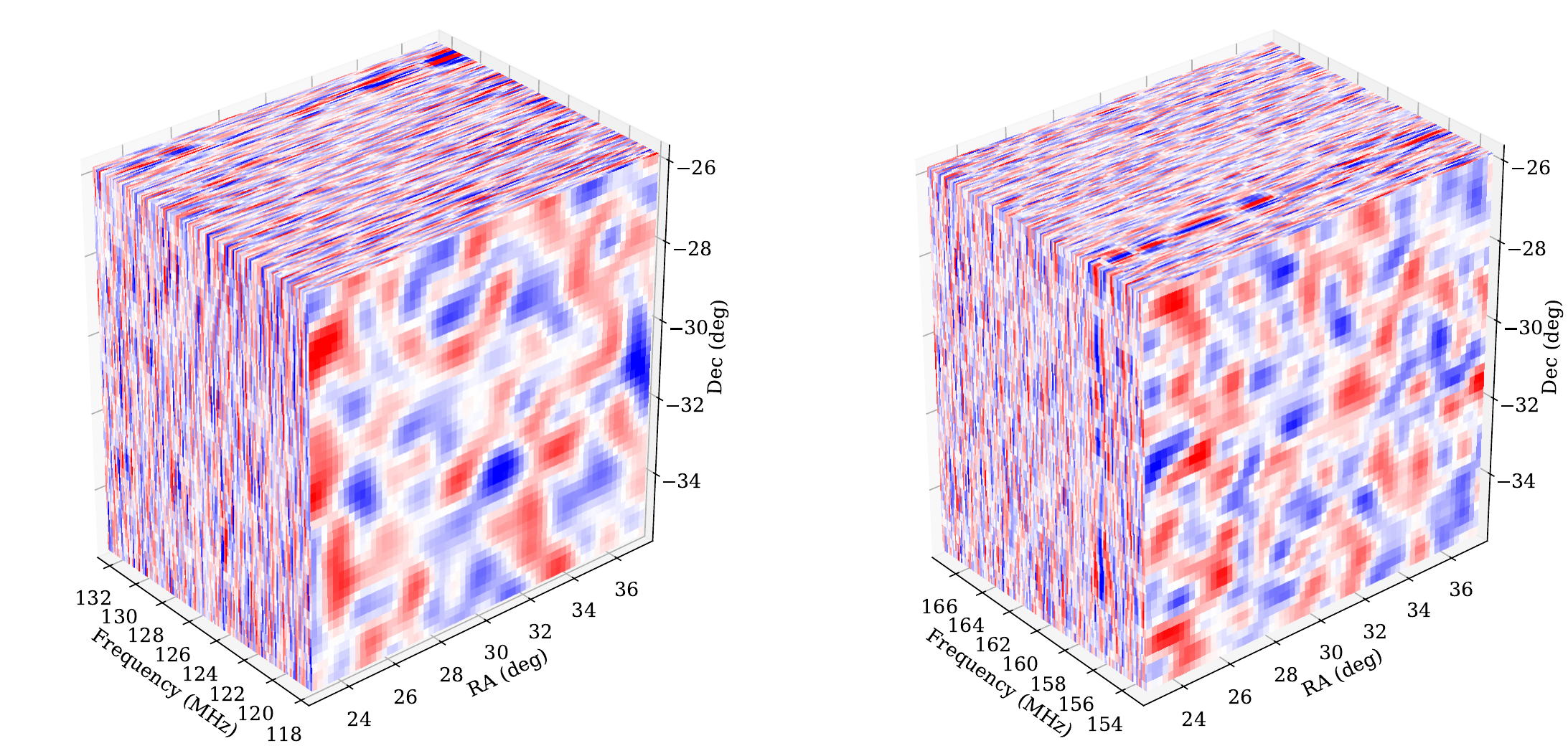}
\caption{Image cubes for wedge-filtered observational data with comprehensive systematics removal. The foreground removal is performed using the \texttt{DAYENU}-filtering method. The wedge-filtered data is expected to be noise-limited. Left and right panels represent Band~1 and Band~2 results, respectively. The maps are centered at 2~hr (Field~I). Those cubes are employed to measure statistical quantities at each frequency or redshift. Bright slices, appearing at 131 and 157~MHz, correspond to systematics residuals discussed in Section~\ref{sec:nongaussianity_H1C}.}
\label{fig:wedge-filtered_cube}
\end{figure*}

Specifically, thermal noise in the visibility data per baseline follows
\begin{align}
    \mathbf{n} \sim \frac{1}{\sqrt{2}}\mathcal{N}(0,\sigma_n^2) + \frac{i}{\sqrt{2}}\mathcal{N}(0,\sigma_n^2),
    \label{eqn:noise_sim}
\end{align}
where $\mathcal{N}(0,\sigma_n^2)$ is a normal distribution with a zero mean and variance specified by Equation~\eqref{eqn:thermal_noise}. We draw samples of the noise visibility based on Equation~\eqref{eqn:noise_sim} using a Monte-Carlo simulation and perform forward modeling to form noise images as follows:
\begin{align}
    \hat{\mathbf{x}}_{\rm N} =  \mathbf{D} \mathbf{A}^\dagger \mathbf{N}^{-1} \mathbf{n}.
    \label{eqn:image_noise}
\end{align}

The noise properties in the image domain can be understood through the map noise covariance matrix, representing the correlated noise between pixels, which can be compute as follows, 
\begin{align}
    \mathbf{C}_{\rm N} = \mathbf{D} \mathbf{A}^\dagger \mathbf{N}^{-1} \mathbf{A} \mathbf{D}^{\rm T} = \mathbf{P} \mathbf{D}^{\rm T}.
    \label{eqn:cov_noise}
\end{align}
The noise variance, diagonal elements of $\mathbf{C}_{\rm N}$, can be explicitly written as,
\begin{align}
    C_{pp} &= P_{pp}D_{pp} \\
    &\approx \frac{\sum_t B_{pt}^2}{\sum_t B_{pt}} \frac{\bar{\sigma}_n^2}{\sum_t B_{pt}} \\
    &= \bar{\sigma}_n^2 \frac{\sum_t B_{pt}^2}{\big(\sum_t B_{pt}\big)^2},
    \label{eqn:noise_variance}
\end{align}
Note that the second equation is derived under the assumption that the noise variance of visibility remains approximately constant over time. For a single time integration, the variance of a pixel is equal to $\sigma_n^2$, which indicates the noise variance is unaffected by the beam and independent of the pixel's location. For longer integration, the noise is attenuated by the primary beam at pixels far away from the zenith. However, because the signals also experience attenuation at these off-zenith pixel locations due to the primary beam, S/N ratio may not see improvement at such pixel locations. For example, a naive estimation of S/N ratio using Equations~\eqref{eqn:P_matrix} and \eqref{eqn:noise_variance} suggest S/N $\propto \sqrt{\sum_t B_{pt}^2}$, indicating the decrease of S/N at pixel locations distant from the beam center.

Mapmaking processes for noise data of the Phase~I observation and future observation are more detailed in Sections~\ref{sec:data_cubes} and \ref{sec:detectability_H320}, respectively.

\section{Wedge-filtered Data Cube Construction for Phase~I Observation}

\label{sec:data_cubes}
In this section, we provide a description of the data products used for measuring one-point statistics for the Phase~I observation. As we are interested in measuring one-point statistics at each redshift, data cubes in the dimension of RA $\times$ Dec $\times$ frequency are constructed for observational and simulation data.

As described in Section~\ref{sec:obs_data}, we use calibrated observational visibility data for Band~1 and Band~2 for each $EE$- and $NN$-polarizations. In order to remove noise biases for variance measurements (note that skewness is an unbiased measurement), we divide the visibility data into even and odd time groups by selecting every other timestamp. If thermal noise in the maps remains uncorrelated across different time groups, it is expected that when cross-multiplying maps from these time groups, the resulting variance will be free from biases introduced by noise. We utilize observational data with and without systematics mitigation to investigate the effect of systematics in measuring one-point statistics in the image domain.

The visibility datasets undergo the \texttt{DAYENU} filtering with the buffer size of 300~ns and the foreground suppression factor of $10^{-9}$. The filtered data are converted into images through the mapmaking process (Section~\ref{sec:dom}) for each frequency channel, polarization, and time group (even and odd). Pseudo stokes-I maps are made by averaging two polarization maps. Image cubes for Band~1 and Band~2 are constructed for each hour of LST, spanning 15$\times$10 square degrees in RA $\times$ Dec, which are then stitched together to cover each field as defined in Section~\ref{sec:obs_data}.

An example for the wedge-filtered data cubes of observational data is presented in Figure~\ref{fig:wedge-filtered_cube}. Left and right panels correspond to the image cube of Band~1 and Band~2, respectively, centered at the 2~hr LST field (Field~I) for a 15$\times$10 square degree field. The cubes are generated from observational data, with comprehensive systematics mitigation, implying that they should exhibit random Gaussian fields from thermal noise. However, even after the systematics were subtracted, some systematics residuals may still be present, as indicated by bright slices appearing at around 131 and 157~MHz. We will discuss statistical measurements on the image cubes in Section~\ref{sec:nongaussianity_H1C}.

HERA observations can achieve higher resolution along the line-of-sight (or frequency) direction, approximately 2~cMpc, compared to the spatial direction, which is around 60~cMpc for the entire HERA core array. The finer resolution along the frequency axis than the spatial direction is illustrated in Figure~\ref{fig:wedge-filtered_cube}.

For the 21~cm simulation data described in Section~\ref{sec:eor_simulations}, we convert the simulated visibility data into data cubes for each band after the \texttt{DAYENU} filtering with the same buffer size and the suppression factor as the observational data. Though the 21~cm simulation is noiseless, we split the visibility data into different time groups to be consistent with the process of the observational data. All four distinct EoR models and all mock visibilities to estimate the sample variance undergo the same process.

As described in Section~\ref{sec:noise_simulations}, thermal noise is realized in visibility space using a Monte-Carlo simulation according to the specification of the Phase~I observation. We simulate 1000 sets of random noise visibilities, spanning a LST range of 0--15~hr for $EE$- and $NN$-polarizations. We divide each visibility data into even and odd timestamps. The visibility data of each time group is then \texttt{DAYENU}-filtered and transformed to pseudo stokes-I maps as the observational data. The noise image cubes are used for statistical inference of observational data compared to expected noise levels in Sections~\ref{sec:cc_mitigation} and \ref{sec:model_comparison}. 

\section{Effects of Systematics on Measuring One-Point Statistics}
\label{sec:cc_mitigation}
\subsection{Systematics Residual in Wedge-Filtered Data}
\label{sec:systematic_P(D)}
Major components of systematics that were mitigated in the HERA Phase~I observation are the cable reflection and crosstalk (\Ha{}, \Hthree{}). The reflection in the coaxial cable may occur due to an impedance mismatch at the termination of the cable. This reflection results in duplicates of signals appearing at a certain delay, equivalent to twice the light travel time along the cable. This systematic effect is addressed through reflection calibration, a process that involves fitting reflection coefficients using the autocorrelation visibility within a delay range of 150--1500~ns \citep{Kern2019, Kern2020b}. This calibration method effectively removes the reflection terms through a direction-independent calibration process.

\begin{figure}[t!]
\centering
\includegraphics[scale=0.4]{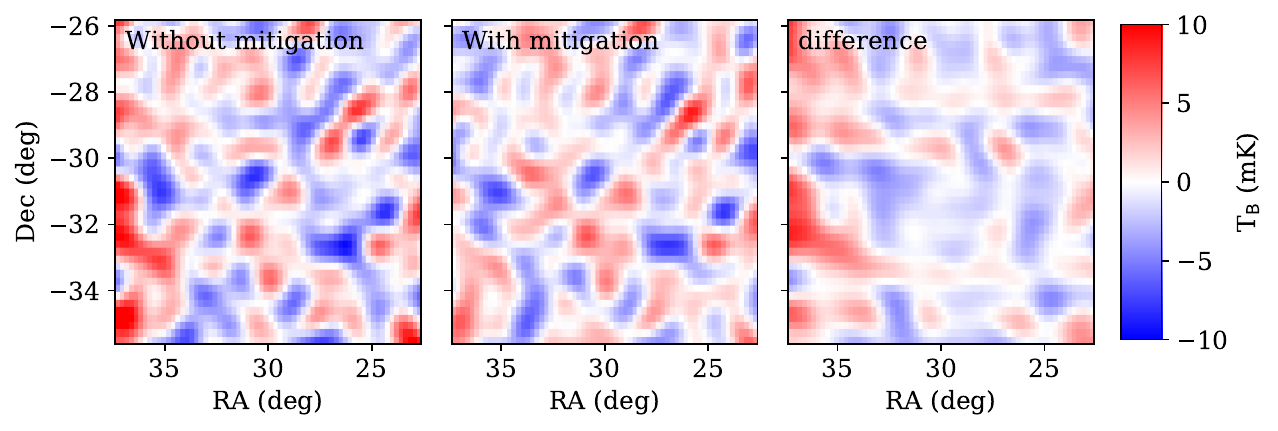}
\caption{Comparison of maps before (first panel) and after systematics mitigation (second panel) at 126.0~MHz in Field~I. The last panel shows the difference between them, which is of a similar order of magnitude as the mitigated result, indicating the data before mitigation is systematics-limited.}
\label{fig:comp_cc_mitigation}
\end{figure}

Crosstalk is another major systematic in Phase~I observations, potentially affecting a wide range of delays. It is hypothesized that feed-to-feed reflections can imprint autocorrelation signals onto cross-correlations at higher delays. Since autocorrelations vary slowly over the beam crossing timescale, a mitigation strategy based on Singular Value Decomposition (SVD) was applied per baseline to isolate time-delay modes contaminated by crosstalk. This was combined with Gaussian Process Regression (GPR) low-pass filtering along the time axis to further suppress systematics \citep{Kern2019, Kern2020b}. The full systematic mitigation used in this study was implemented by \Hthree{}.

Figure~\ref{fig:comp_cc_mitigation} exhibits the wedge-filtered maps without and with the systematics mitigation described above. Notably, certain prominent diffuse structures evident in the first panel are no longer present after the mitigation process (second panel). In the absence of appropriate mitigation, wedge-filtered maps might retain foregrounds introduced by systematic effects. The last column, illustrating the difference between the two maps, potentially indicates the presence of foreground residuals leaking into high delay modes in the absence of systematics mitigation.

\begin{figure}[t!]
\centering
\includegraphics[scale=0.45]{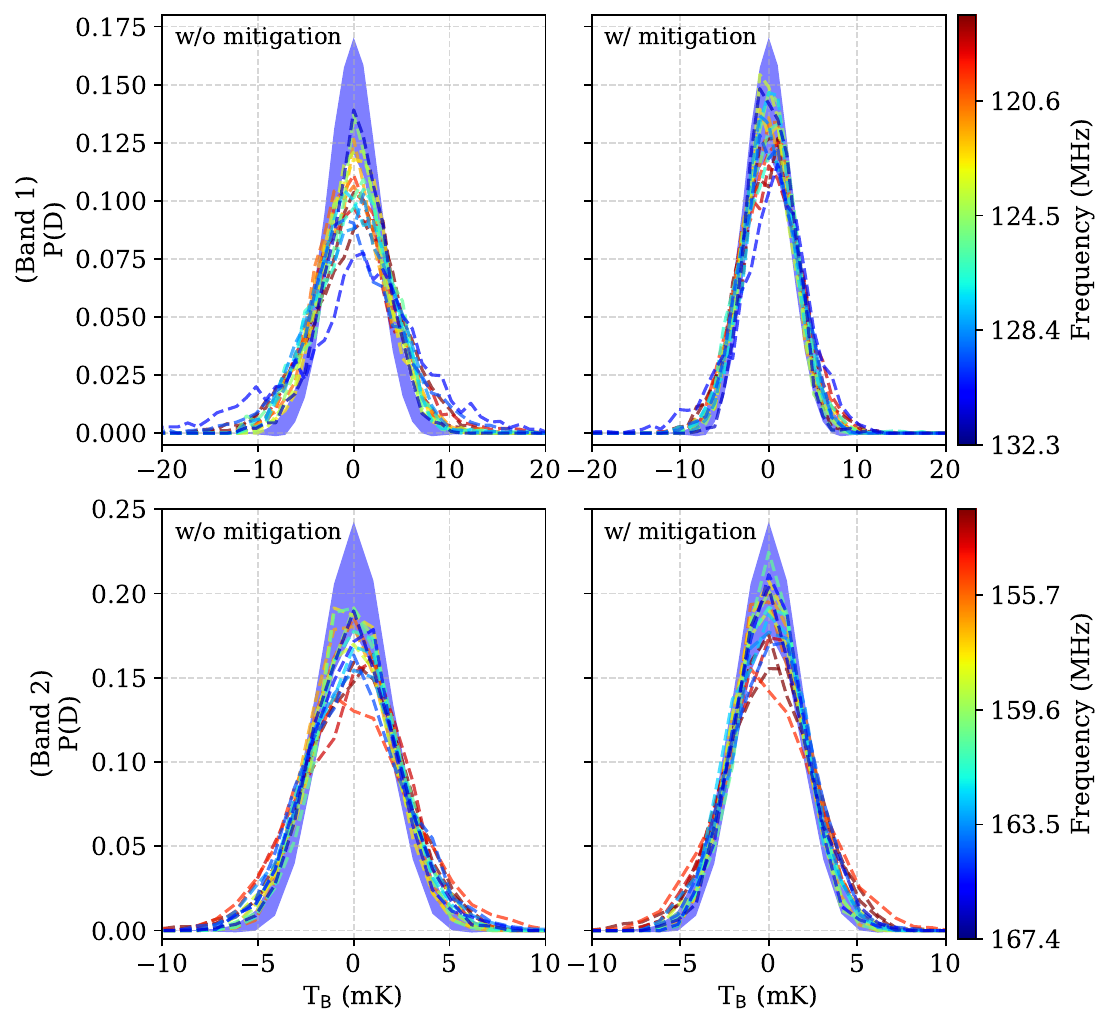}
\caption{P(D) of wedge-filtered maps for data before the systematics mitigation (left column) and after the mitigation (right column) for Band~1 (top) and Band~2 (bottom) in Field~I. The 95\% confidence interval of noise simulation distributions is presented as a blue shaded region for reference. The deviation in P(D) from the reference distribution for the data without systematics mitigation is reduced after applying the systematics mitigation, although some outliers still remain.}
\label{fig:P_D_cc}
\end{figure}

To investigate the effects of systematics on map properties, we employ a P(D) analysis \citep{Scheuer1957, Condon1974}. P(D) represents a density profile of pixel values in brightness temperature, where random thermal noise is expected to follow a Gaussian distribution. The results for Field~I are shown in Figure~\ref{fig:P_D_cc}. The top and bottom panels correspond to Band 1 and Band 2, respectively. Every 10th channel from each band is selected for illustration purposes. Maps from even and odd time groups are averaged. To examine how systematics affect the distribution shape, P(D) is measured from maps of the wedge-removed data before (left) and after (right) systematic mitigation. The 95\% confidence intervals of 1000 noise simulations are shown as blue shaded regions for reference.

In the absence of systematics removal, for most frequency channels, noticeable deviations from a Gaussian profile, such as broadening, asymmetry, and tailedness, are evident in the P(D), indicative of non-Gaussianity caused by systematics. After systematics subtraction in the HERA Phase~I observation, where the data is expected to be noise-limited, the P(D) for most channels are aligned with the expected noise profile, but there are still some outliers. We further investigate the effects of systematics residuals on one-point statistics measurements quantitatively over a range of frequency in the following section.

While systematics mitigation for cable reflections and crosstalk has been shown to have minimal effects on power spectrum estimation \citep{Kern2019, Kern2020b}, its impact on one-point statistics of the EoR signals remains unexplored in this study. Systematics mitigation may introduce correlations in the data, necessitating more accurate noise characterization that accounts for potential correlated noise properties in one-point statistical analyses. A comprehensive investigation into these effects is left for future work.

\begin{figure*}[t!]
\centering
\includegraphics[scale=0.7]{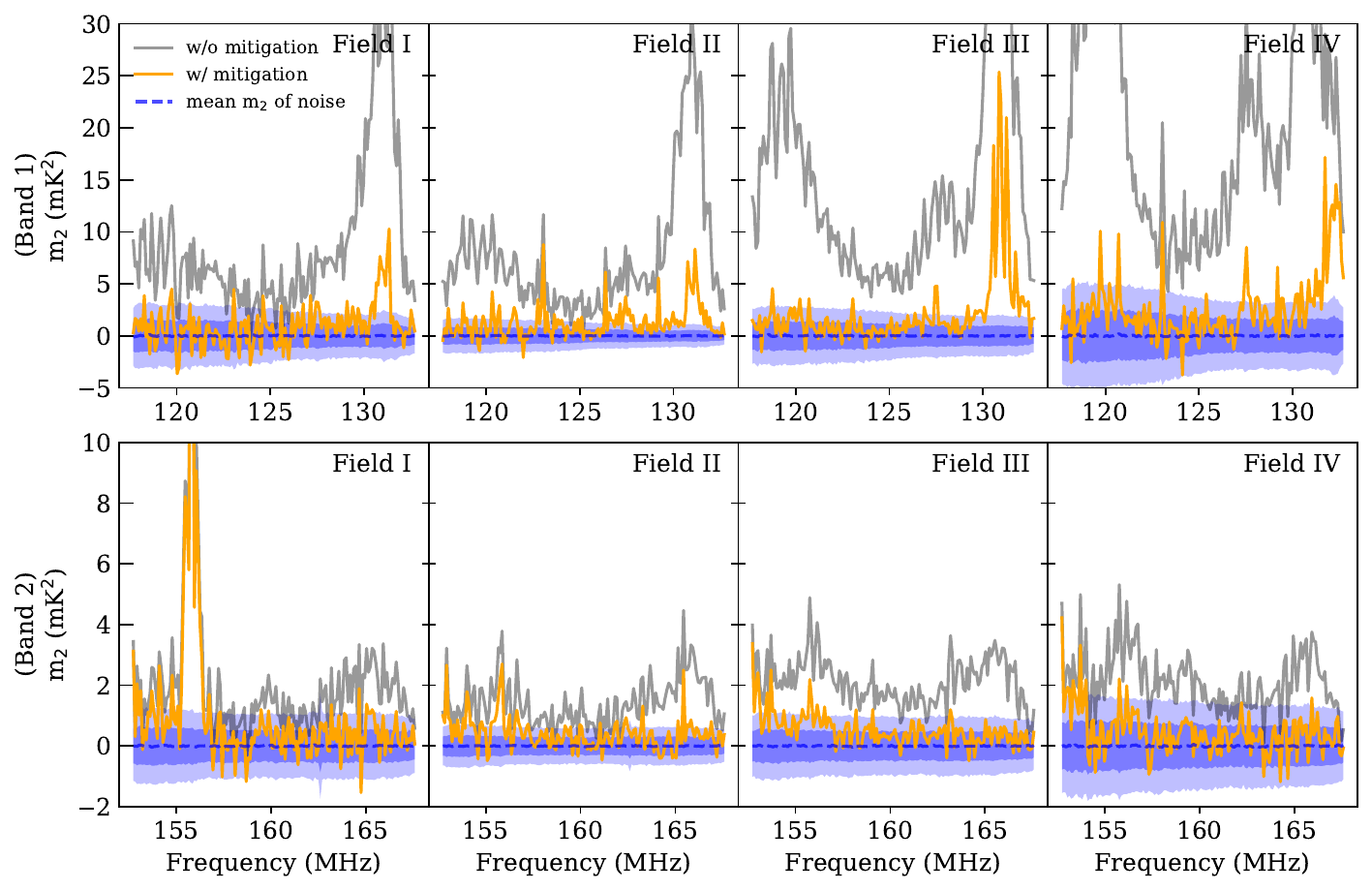}
\caption{$m_2$, variance measurements, of wedge-filtered observation and noise maps as a function of frequency at different fields. Gray and orange curves represent the data before and after systematics mitigation, respectively. The dark and light blue shaded regions, corresponding to 1$\sigma$ and 2$\sigma$ confidence, respectively, are defined by the 1000 random noise realizations. Both Band~1 (top row) and Band~2 (bottom row) results show the variance is significantly reduced with the systematic mitigation though there are still systematics observed beyond the theoretical noise expectation even after the systematics mitigation for all LSTs.}
\label{fig:variance_w_wo_cc}
\end{figure*}

\subsection{Statistical Measurements for Wedge-removed Observational Data}
\label{sec:nongaussianity_H1C}
As observed in Section~\ref{sec:systematic_P(D)}, some residual systematics remain in the wedge-filtered map even after the mitigation efforts. In this section, we examine the systematics effects on one-point statistics of observational data in comparison to expected theoretical noise across a range of frequencies.

Variance is a primary quantity in one-point statistics which is related to the power spectrum,
\begin{align}
    \sigma^2 = \int\frac{d^3 k}{(2\pi)^3} P(\mathbf{k)},
\end{align}
where $P(\mathbf{k})$ is a power spectrum of signals. If there are systematic residuals in the data, particularly foreground residuals leaking into the EoR window in the power spectrum, the variance tends to increase and becomes sensitive to the presence of these systematic residuals.

To eliminate noise bias in variance measurements, we cross-multiply the even and odd wedge-removed maps and average the product across pixels at each frequency channel for both observational data and noise simulations for each field.

The variance measurements, or $m_2$, are presented in Figure~\ref{fig:variance_w_wo_cc} for Band~1 (top) and Band~2 (bottom) at each field. The y-axis represents the variance after removing the noise bias. The dashed line represents the mean, while the dark and light blue shaded regions indicate the 1$\sigma$ and 2$\sigma$ confidence intervals, respectively, based on 1000 noise simulations. The noise simulation fluctuates around a zero mean, as expected. A smaller root-mean-square (RMS) of the $m_2$ measurements is achieved in Field II, as it is a relatively quiet radio sky in terms of foreground noise.

In both Band~1 and Band~2, there are substantial excess of variance in the presence of systematics (gray curve) as expected. The oscillating patterns have a period of approximately 1 MHz, corresponding to 1000 ns, which is a characteristic delay of crosstalk. After implementing systematics mitigation, the excess of variance is reduced, with a more pronounced effect observed in Band 1. However, the orange curves still show variance excess, indicating remaining systematics, especially more prominent in Field II with higher sensitivity observations. This excess of variance could arise from systematics residuals of cable reflections, crosstalk, faint RFIs, and/or chromatic gain errors introducing foreground leakage into high delay modes (\Ha{}, \Hthree{}).

\begin{figure*}[t!]
\centering
\includegraphics[scale=0.7]{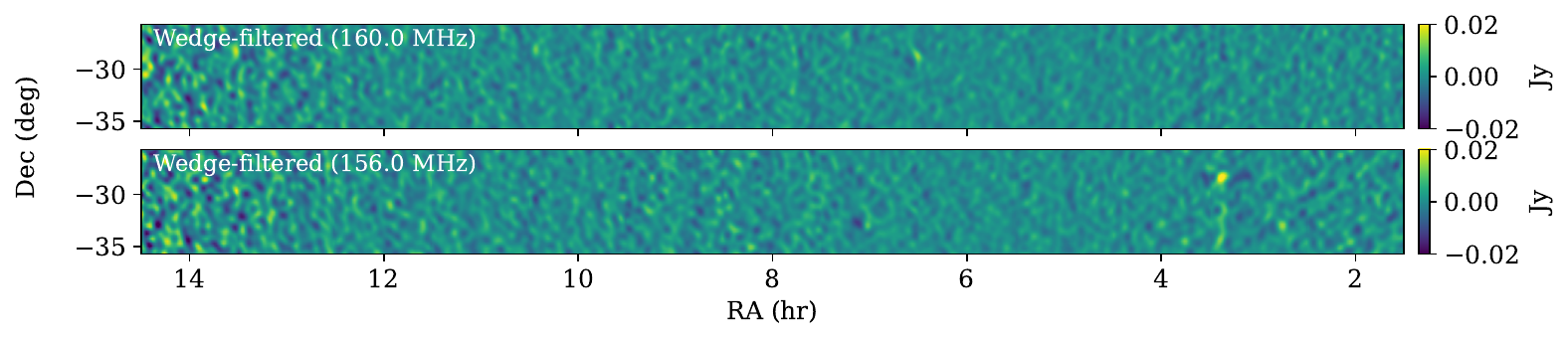}
\caption{HERA stripe for wedge-removed observational data. The top and bottom panels display maps at 160.0 MHz and 156.0 MHz, respectively. The top panel shows a map that is relatively free of systematics, whereas the bottom panel exhibits a distinct foreground residual feature at 3.3 hr, which are residuals of the grating lobes of Fornax~A.}
\label{fig:hera_stripe_wedge_filtered}
\end{figure*}

The removal of strong and broad RFIs around 137~MHz (associated with ORBCOMM satellites) and 150~MHz during calibration results in significant discontinuities in the spectrum (\Ha{} \Hthree{}). These gaps may interact with the data reduction process, including systematics mitigation, inpainting of removed data points, and smoothing of calibration gains \citep[e.g.,][]{Chen2025}. This interaction might explain unsuccessful removal of systematic residuals including the observed peaks around 131~MHz and 156~MHz. Although these features are attenuated in power spectrum estimation due to a tapering function, they can be manifest in our imaging analysis.

Figure~\ref{fig:hera_stripe_wedge_filtered} illustrates an example of systematics residuals, or foreground residuals, in image space after systematics mitigation. Wedge-removed maps at different frequency channel are presented. Each panel has the same dimension of Figure~\ref{fig:hera_stripe_foreground}. At 160.0~MHz (top panel), the foreground-removed map displays a relatively clean map with minimal foreground contamination. In contrast, at 156.0~MHz (bottom panel), there are noticeable features of foreground residuals leaking into the EoR window at 3.3~hr, corresponding to the residuals of grating lobes from Fornax~A. We find the presence of foreground residuals across a range of frequencies, which may explain the variance excess shown in Figure~\ref{fig:variance_w_wo_cc}. The increasing amplitude of noise at around 14~hr in both maps is due to the sky approaching the bright galactic center.

\begin{figure*}[t!]
\centering
\includegraphics[scale=0.7]{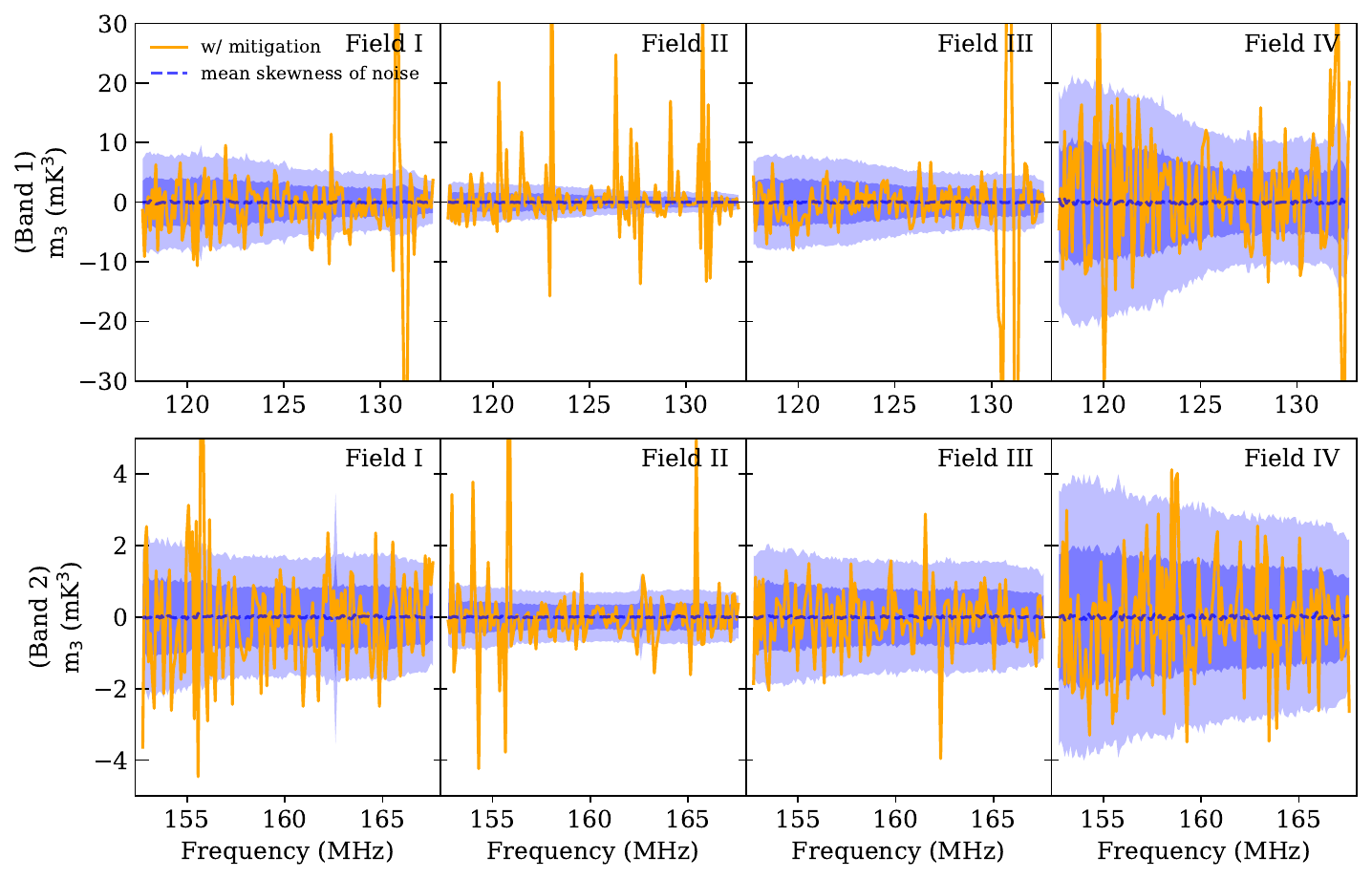}
\caption{$m_3$ measured from the wedge-filtered observation, along with the noise simulations denoted as the mean, 1$\sigma$, and 2$\sigma$ confidence intervals, at different fields. Orange curves represent the data with systematics mitigation. Though there are some outliers observed beyond the 2$\sigma$ boundary, $m_3$ measurements for large fraction of frequency channels are confined within the 2$\sigma$ range except for Field~II.}
\label{fig:skewness_w_wo_cc}
\end{figure*}

To separate the behaviors of $m_2$ and $m_3$, instead of the standardized third moment $S_3$, we use the $m_3$ as a statistical indicator of non-Gaussianity. As evident in Figure~\ref{fig:P_D_cc}, some frequency channels affected by systematics exhibit pronounced non-Gaussian features, even with the mitigation efforts. In Figure~\ref{fig:skewness_w_wo_cc}, we depict the $m_3$ as a function of frequency for Band~1 and Band~2 in each field. Because $m_3$ is an unbiased measurement with respect to the noise, the maps of even and odd time groups are averaged before we measure the $m_3$. The blue dashed lines represent the mean, and the dark and light shaded regions denote the 1$\sigma$ and 2$\sigma$ confidence intervals of $m_3$ respectively, as calculated from 1000 noise simulations. As expected the thermal noise drawn from a random Gaussian distribution has a zero mean skewness.

The results for the data after the systematics mitigation are shown in orange lines. The $m_3$ measurements oscillate around the zero mean, which is largely consistent with the noise simulations. However, there are peaks surpassing the 2$\sigma$ boundary across all fields, particularly noticeable in Field~II as we see in Figure~\ref{fig:variance_w_wo_cc}. Unlike the variance, systematics effects in $m_3$ can be either positive or negative values. The results from $m_2$ and $m_3$ underscore the significance of mitigating systematics to a deeper level for exploring one-point statistics of 21 cm signals, especially in future observations with HERA.

\section{Comparison with 21~cm Simulations}
\label{sec:model_comparison}
As detailed in Section~\ref{sec:eor_simulations}, we simulate four distinct EoR models to investigate one-point statistics measurements to assess their detectability. In this section, we perform a likelihood analysis for $m_2$ measurements and explore the detectability of $m_3$ measurements based on Phase~I observations compared to the 21~cm model.

Previous studies adopted analytic calculations to estimate the errors associated with the statistical quantities \citep[e.g.,][]{Watkinson2014, Kittiwisit2017}. In this study, rather than relying on an analytic form, we employ 1000 noise Monte Carlo simulations with a brute-force approach. This method allows us to estimate the uncertainties of the mean of statistical quantities and directly calculate the error bars on the observational data. It accounts for correlations between pixels induced by instrumental effects and correlations between frequency channels introduced by wedge-filtering.

\begin{figure*}[t!]
\centering
\includegraphics[scale=0.7]{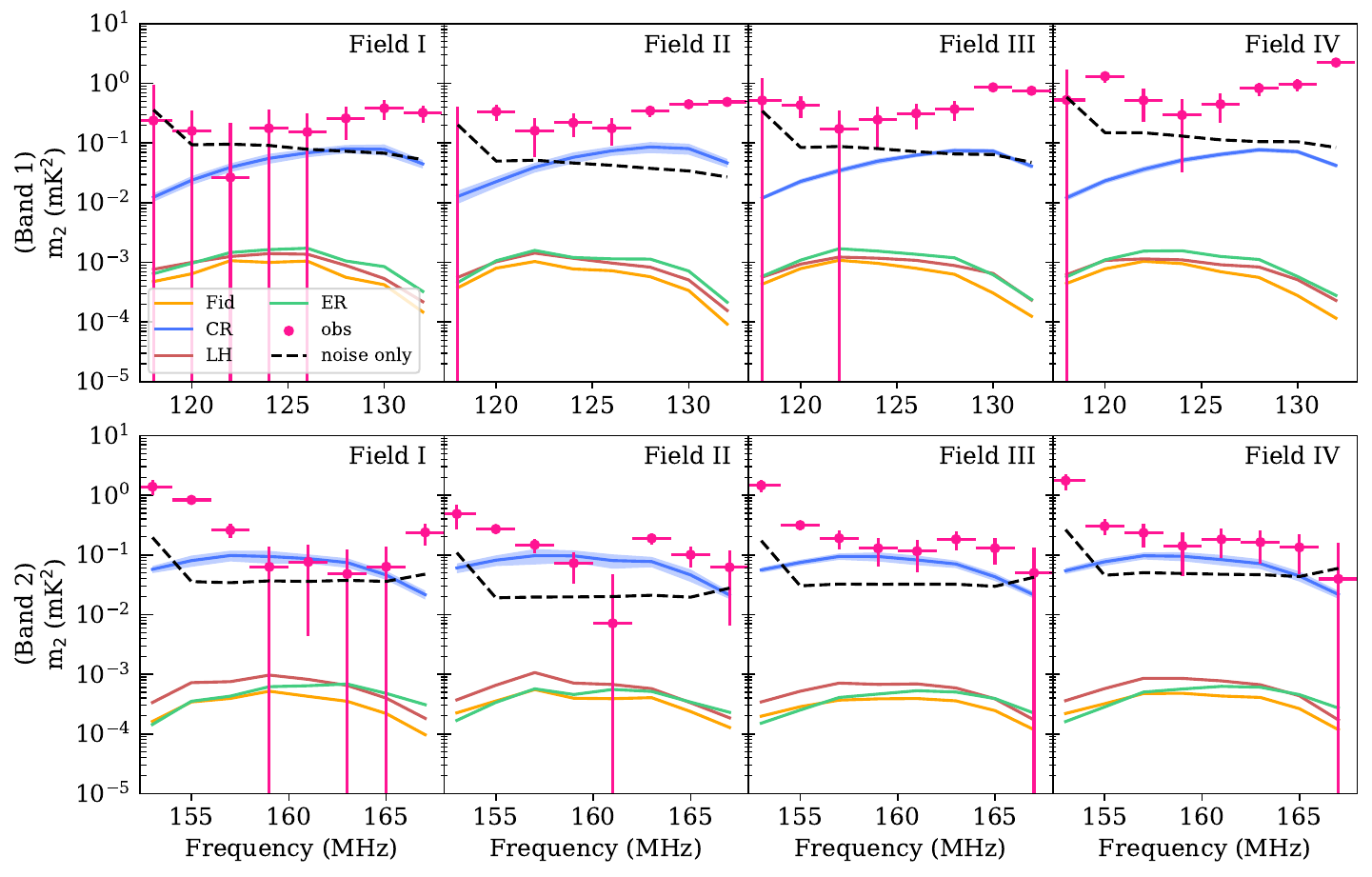}
\caption{$m_2$ measurements based on wedge-filtered data in comparison to simulations for each field and each band. Frequency averaging over 0.5~MHz bin is performed to improve the sensitivity. Variance is measured from every 2~MHz sliced volume. The 1$\sigma$ errors computed from noise simulations are indicated by the black dashed lines, providing the theoretical expectation. The $m_2$ values of the observational data are represented by pink symbols with 2$\sigma$ error bars, derived from the same noise simulations, along the y-axis. The x-axis error bar represents the bin width. The lowest upper limit, considering the 2$\sigma$ error bar, is 0.047~mK$^2$ at 161~MHz in Field~II. Additionally, four 21~cm models, including the fiducial (Fid), cold reionization (CR), large halos (LH), and extended reionization (ER) models, are compared to the observational data points. These simulations do not include thermal noise. Blue shaded regions correspond to the sample variance of the cold reionization model derived from 30 samples.}
\label{fig:variance_detectability}
\end{figure*}

\subsection{Statistical Tests on $m_2$}
\label{sec:stat_test_m2}
While variance does not measure non-Gaussianity, it is useful for evaluating the likelihood of an EoR model based on the fluctuations of 21~cm signal of each model relative to observational data, analogous to the power spectrum analysis. Figure~\ref{fig:variance_detectability} presents the $m_2$ measurements for the four 21~cm models in comparison to the observational data points in each field. We use the dataset with systematics mitigated. To increase sensitivity in our observations, we coherently average the intensity maps over every 5 channels, roughly corresponding to a 0.5~MHz bin size, which improves sensitivity in $m_2$ by about 5 times. We then bin the image cube into sub-band data volumes using a 2~MHz interval between 117--133~MHz for Band~1 and between 152--168~MHz for Band~2. The variance is measured in each sub-band volume, further enhancing sensitivity in $m_2$ by about 2 times. As in the previous section, we cross-multiply image volumes from even and odd time groups to remove noise bias. The same process is also applied to the maps of wedge-filtered noise and 21 cm simulations.

We use the noise simulations to estimate the RMS of the expected noise, represented by black dashed lines in Figure~\ref{fig:variance_detectability}, which provides the theoretical expectation when the foreground-removed observational data is noise-dominated. This corresponds to dark shaded region in Figure~\ref{fig:variance_w_wo_cc} but with deeper noise level as we average intensity maps and measure the variance in a volume as described above. Additionally, the noise simulations are used to estimate the observational error bars for each sub-band, without accounting for the effects of systematics on these error bars.

We show the $m_2$ measurements for Band~1 (top row) and Band~2 (bottom row) in Figure~\ref{fig:variance_detectability}, where the observational data is denoted as pink circles with 2$\sigma$ error bars from the noise simulations. The data points with $m_2$ smaller than zero are set to zero and the upper limit is set by the 2$\sigma$ error bars from the zero. The data points with positive values have a upper limit of $m_2+2\sigma$. In addition, we illustrate the $m_2$ measurements for 21~cm simulation data including fiducial, cold reionization, large halos, and extended reionization models. Sample variance is illustrated as a shaded region for the cold reionization model, accounting for roughly 15\% of the amplitude of $m_2$.

It is apparent that some frequency bins are heavily contaminated by systematics residuals, leading to a large variance beyond the 2$\sigma$ error bar compared to the expectation. The variance excess is more evident across frequency bins than we see in Figure~\ref{fig:variance_w_wo_cc} because the thermal noise is integrated down through averaging the images along the frequency while the systematics may not be integrated down. \citet{Morales2018} studied the effects of calibration errors on an imaging power spectrum in comparison with a delay-spectrum power spectrum. They found that the imaging power spectrum is more sensitive to frequency-independent calibration errors than the delay-spectrum power spectrum because calibration errors from different baselines introduce spectral structure in the image space. The integral of the imaging power spectrum corresponds to the variance we measure, which may explain why our variance measurements are sensitive to systematics.

If the data is free from systematics and aligned with the thermal noise expectation, Band~2 may be employed to reject the cold reionization, especially in Field~II as the data points are lower than the model. However, in the presence of the systematics residuals, it is not trivial to make such conclusion. For Band~2, there is one frequency bin in Field~II where the $m_2$ value of the observational data, when considering the 2$\sigma$ error bar, is lower than that of the cold reionization model. The upper limits could help constrain certain model parameters.

\begin{figure}[t!]
\centering
\includegraphics[scale=0.65]{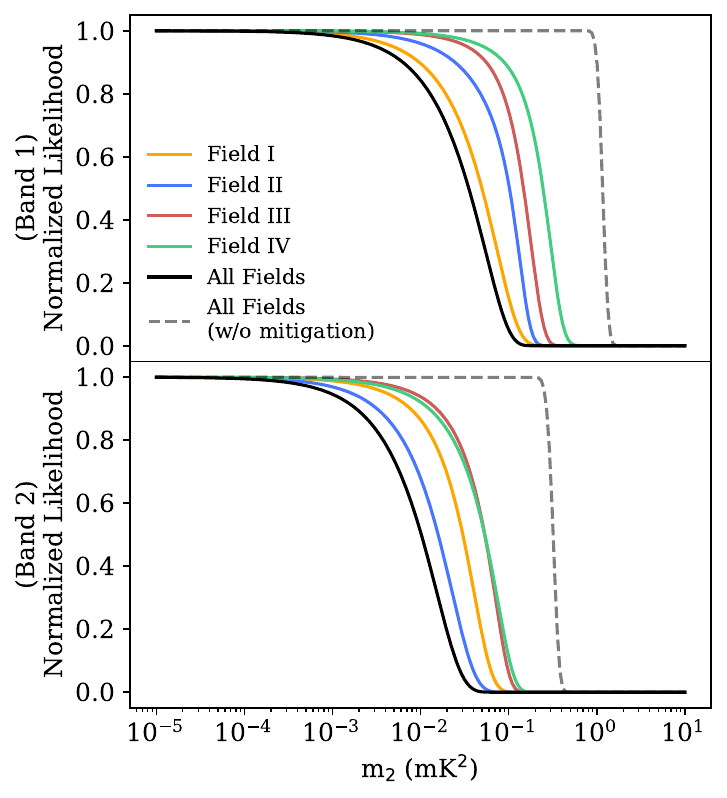}
\caption{Normalized likelihood for $m_2$, marginalized over systematics defined in Equation~\eqref{eqn:marginalized_likelihood_m2}. We present the normalized likelihood for each field for Band~1 (top) and Band~2 (bottom), using all observational data points shown in Figure~\ref{fig:variance_detectability}. The black solid line is the likelihood considering all data points across all fields. To see the impact of systematics mitigation, we provide the likelihood without the mitigation, based on all fields (gray dashed line). We assume the variance model of $m_2 \propto \nu^0$. The cold reionization model has $m_2$ about 0.07 mK$^2$ on average in Band~2, which implies the model is likely to be rejected by the likelihood analysis.}
\label{fig:m2_likelihood}
\end{figure}

To understand the impact of the variance measurements on evaluating 21~cm models, we employ a likelihood analysis, similar to the approach used in \Hb{} and \Hthree{}. The likelihood is a probability of the observed data given different values of the model parameters. Assuming random Gaussian thermal noise, the likelihood is written as,
\begin{align}
    \mathcal{L}(\mathbf{d}|\mathbf{\theta}, \mathcal{M}, \mathbf{u}) \propto \exp{\bigg(-\frac{1}{2}\mathbf{r}(\mathbf{\theta}, \mathbf{u})^{\rm T} \Sigma^{-1} \mathbf{r}(\mathbf{\theta}, \mathbf{u})}\bigg),
    \label{eqn:likelihood_m2}
\end{align}
where $\mathbf{r(\theta, \mathbf{u}}) = \mathbf{d} - \mathbf{u} - \mathbf{W}\mathbf{m}(\theta)$, $\mathbf{d}$ is the $m_2$ measurement of observational data, $\theta$ are parameters of a model $\mathcal{M}$, and $\mathbf{u}$ is the systematics. $\mathbf{W}$ is a window function that maps a theoretical model to observable. Because we forward model our measurements, our 21~cm measurements account for the effect of the window function. $\Sigma$ represents a covariance matrix of the observational data and models.

Assuming that the sky signal is much smaller than the systematics, one can show that the effect of the systematics is only to add (and not subtract) power to the variance \citep{Morales2023}. This is supported by Figure~\ref{fig:variance_w_wo_cc}.
Given the assumption, a marginalized likelihood over systematics can be obtained by integrating Equation~\eqref{eqn:likelihood_m2} with the constraint $\mathbf{u} \geq 0$,
\begin{align}
    \mathcal{L}(\mathbf{d}|\mathbf{\theta}, \mathcal{M}) \propto \prod^{N_{\rm d}}_i \frac{1}{2}\bigg(1 + {\rm erf}\bigg[\frac{t_i}{\sqrt{2} \sigma_i}\bigg]\bigg),
    \label{eqn:marginalized_likelihood_m2}
\end{align}
where ${\rm erf}$ is the error function, $t_i$ is $i$th element of $\mathbf{t} = \mathbf{d} - \mathbf{W}\mathbf{m}$, and $N_{\rm d}$ is the number of frequency bin. $\sigma_i$ is the square root of the quadrature sum of the variance derived from noise simulations and the sample variance corresponding to 15\% of the model variance. In this derivation, we assume each bin is uncorrelated. The details on the derivation are described in \Hb{}.

In Figure~\ref{fig:m2_likelihood}, we displays the normalized likelihood, computed for all data points of each field given in Figure~\ref{fig:variance_detectability}. The ``All Fields'' result (black solid line) is based on all data points from all four fields. Top and bottom panels represent Band~1 and Band~2 results, respectively. We assume the model variance of $m_2 \propto \nu^0$ for both bands. The shape of the normalized likelihood serves as an upper limit to constrain feasible model variance. We also present the normalized likelihood before systematics mitigation, indicating a significant improvement after the removal of systematics.

The likelihood for ``All Fields'' in Band~2 is the most effective in constraining models. The cold reionization model has about 0.07 mK$^2$ in $m_2$ averaged over the frequency band of Band~2. The likelihood at $m_2 \sim 0.07~$mK$^2$ is $< 10^{-5}$, suggesting that the cold reionization model is disfavored, supporting findings from power spectrum analyses in \Hb{} and \Hthree{}. To investigate other models, we need an improvement in variance by at least two orders of magnitude.

\begin{figure*}[t!]
\centering
\includegraphics[scale=0.65]{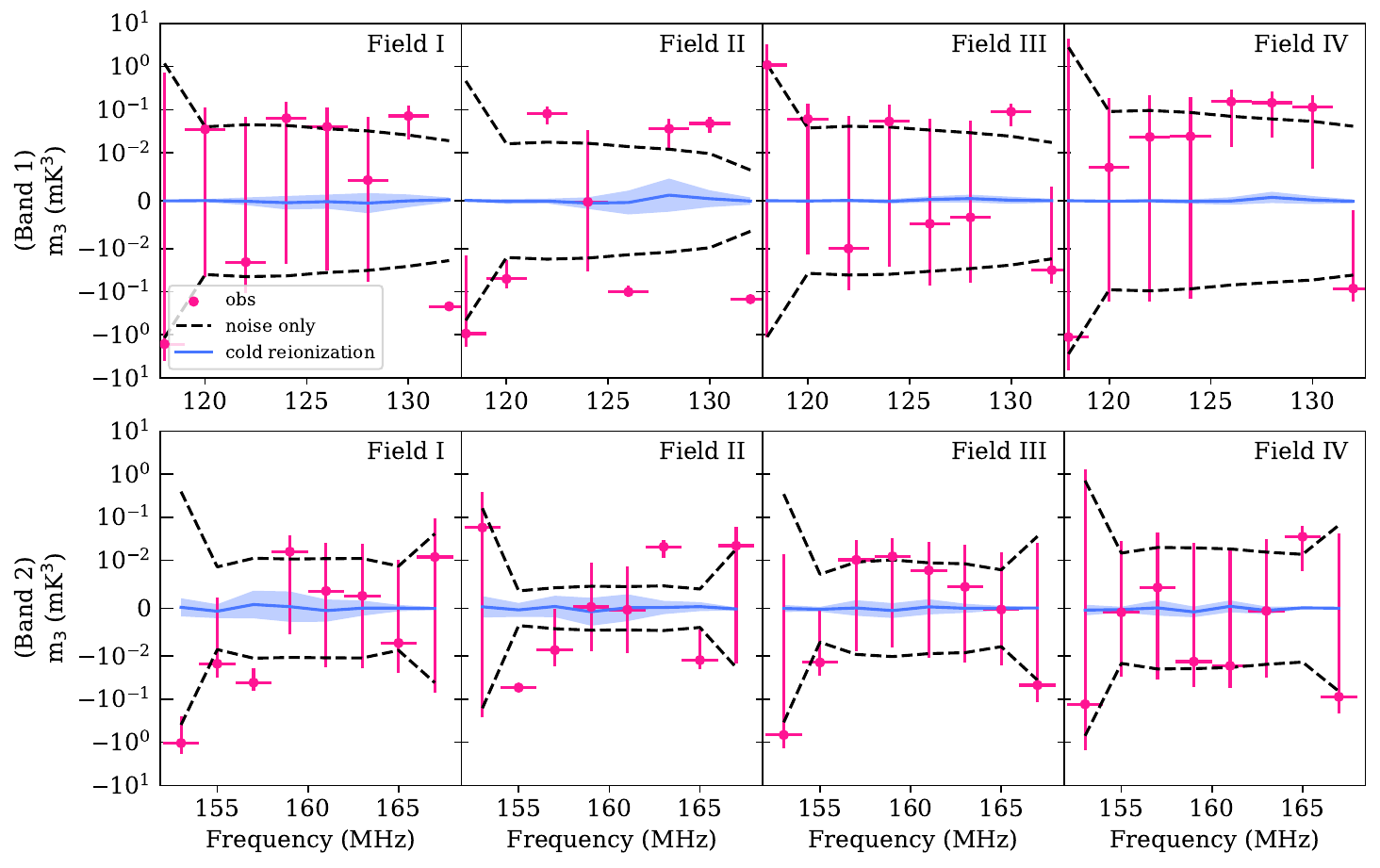}
\caption{$m_3$ measurements for wedge-removed observational data (pink circles) in comparison to simulations for Band~1 (top) and Band~2 (bottom) across all fields. The 1$\sigma$ errors in $m_3$ derived from noise simulations are represented by black dashed lines. The observational data is along with 2$\sigma$ error bars, which are derived from the same noise simulations. Though there are some outliers especially in Field~II, the observational data is largely consistent with the noise simulations. Additionally, the noiseless cold reionization model, characterized by larger variance than other 21~cm models, is depicted with blue lines including the sample variance shown in shaded regions. Due to the instrument resolution and wedge-filtering, the model effectively shows zero $m_3$ as discussed in Figure~\ref{fig:1pnt_psf_filtering_effect_cold_reionization}.}
\label{fig:skewness_detectability}
\end{figure*}

\subsection{Comparison with 21~cm Model based on $m_3$}
\label{sec:model_test_m3}
Investigating EoR models through skewness measurements, considering the sensitivity of the observational data, instrumental effects, and signal loss due to foreground removal, is quite challenging. In this section, we show the $m_3$ measurements of the observational data in comparison with the cold reionization model, which exhibits a larger variance than other 21~cm models.

Similar to Section~\ref{sec:nongaussianity_H1C}, we make image cubes by averaging maps from even and odd timestamps. We average the images for 5 channels and bin the frequency-averaged data into 2~MHz sliced volumes to measure the $m_3$ as we do for the $m_2$ measurements.

We show $m_3$ measurements of wedge-filtered observational data, along with the noise simulations and the noiseless cold reionization model in Figure~\ref{fig:skewness_detectability}. The black dashed lines are the 1$\sigma$ error derived from the noise simulations. Unlike the variance, skewness measurements of 21~cm model can be either positive or negative, and thus we present $m_3$ values of noise simulations in both axes.

The observational data, represented by pink symbols, is closely aligned with the black dashed lines within the 2$\sigma$ error bars, which are also derived from the noise simulations. There are still some frequency bins that are heavily affected by systematics, particularly in Field~II of Band~1, but the overall features of $m_3$ are less affected by the systematics compared to the $m_2$ results shown in Figure~\ref{fig:variance_detectability}. This could be due to the fact that the systematics at each frequency channel can be either positive or negative in $m_3$, and as a result, the effects of these systematics may be smeared out when measuring $m_3$ in 2 MHz volumes. But as we go deeper in our observation, the systematics will be crucial in evaluating 21~cm models using $m_3$ and the systematics should be mitigated in a deeper level.

The $m_3$ of the cold reionization model after the wedge-filtering becomes close to 0, which is shown in blue lines in Figure~\ref{fig:skewness_detectability}. Given the amplitude of the signals and the sensitivity of the observation, drawing a meaningful conclusion from this result is challenging. Furthermore, because the systematics can be either positive or negative, the likelihood defined in Equation~\eqref{eqn:marginalized_likelihood_m2} cannot be used and it is hard to use the likelihood analysis in constraining the model parameters without any constraints on the systematics.

The reduction of $m_3$ values in 21~cm models is partially because the convolution of the instrument response (PSF) tends to wash out small-scale features responsible for non-Gaussianity \citep{Wyithe2007}. An additional decrease arises from the wedge-filtering, expected for the removal of large-scale frequency structures corresponding to small delay modes \citep{Harker2009, Kittiwisit2022}, as discussed in Section~\ref{sec:instrument_effects_1pnt_stat}.  In the next section, we examine the prediction of $m_2$, $m_3$, and $S_3$ measurements assuming full HERA core array.

\begin{figure*}[t!]
\centering
\includegraphics[scale=0.35]{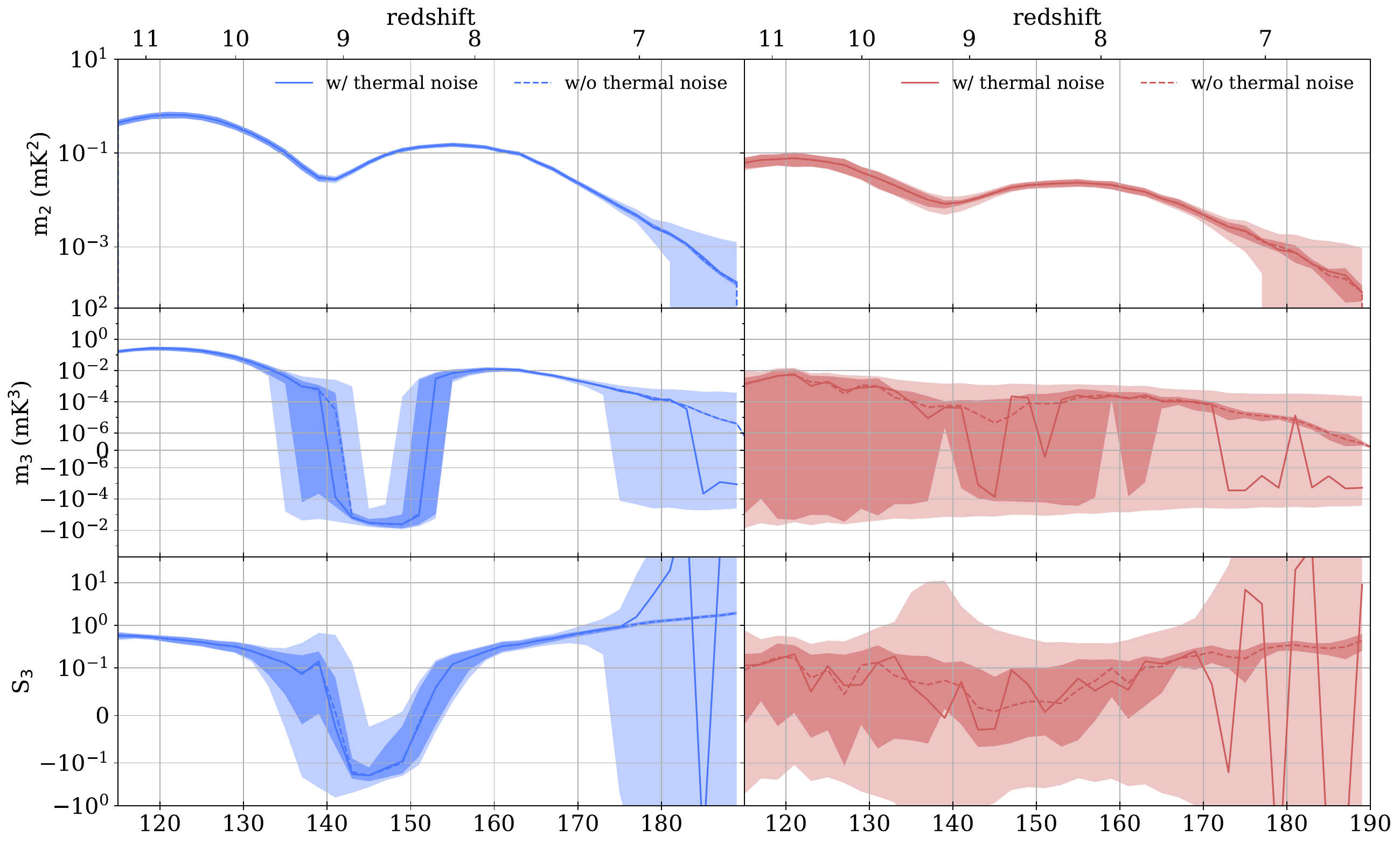}
\caption{$m_2$, $m_3$, and $S_3$ for the fiducial model as a function of frequency with HERA-320, assuming no foreground (left panels) and after applying wedge-filtering (right panels). The dashed lines, along with the dark shaded regions, represent the mean and sample variance, respectively, for the scenario without thermal noise. When thermal noise is included, in addition to the sample variance, the overall uncertainty increases, as indicated by the solid lines and light shaded regions. With wedge-filtering applied, detecting non-Gaussianity becomes more challenging, as shown in the right panels. However, in the absence of thermal noise, there is still potential to observe nonzero $m_3$ at low redshifts. When thermal noise is included (red solid lines and light shaded regions), contamination at low redshifts further obscures non-Gaussian features, making detection in $m_3$ and $S_3$ more difficult. This result could be enhanced by incorporating longer observation durations, utilizing outrigger antennas, exploring alternative EoR models, and applying more advanced foreground removal techniques. For reference, in this model, the midpoint of reionization occurs around $z \approx 8.5$. At $z = 7.3$, where the fractional errors become significant, the ionization fraction reaches 0.95.}
\label{fig:forecast_1pnt_stat_h320}
\end{figure*}

\section{Forecasting Future Observations for One-point Statistics}
\label{sec:detectability_H320}
Previous sections discuss the results from HERA Phase I observations, which utilized a subset of the full array and dipole feeds. The ongoing and upcoming HERA Phase II observations feature a more complete antenna array and a wider frequency coverage with Vivaldi feeds.

As we discuss in Section~\ref{sec:instrument_effects_1pnt_stat} and \ref{sec:model_comparison}, the relatively poor image resolution of Phase~I observations tends to remove the non-Gaussianity information arising from small-scale structure. We extend the analysis to forecast $m_2$, $m_3$ and $S_3$ measurements considering a better resolution of the instrument for future observations using the full HERA core array. The full HERA core array \citep[HERA-320;][]{Dillon2016} consists of 320 antennas with approximately 0.4$^\circ$ resolution. We simulate mock observations using the array configuration, with the fiducial 21~cm model described in Section~\ref{sec:eor_simulations}. A Vivaldi feed beam \citep{Fagnoni2021} is used for the predicted observation.

We consider about 4-hr observation per night, spanning 1.5---5.5~hr in LST at a cadence of 30~seconds. The mock observation is simulated across 115.04--189.99~MHz with a frequency channel width of 244.1~kHz, which is double of the width of HERA Phase~II observation to reduce computation cost. The mock visibilities are wedge-filtered and transformed into images using the same method applied to HERA Phase~I observations. For the wedge-filtering, we consider an optimistic scenario where the foreground is well confined in the foreground wedge with buffer = 0~ns. In addition, we include the perfect foreground removal case for comparison. We repeat the processes 30 times by choosing different sky positions to estimate the sample variance in $m_2$, $m_3$, and $S_3$ measurements.

For the noise estimation, we utilize the system temperature, $T_{\rm sys} = T_{\rm rec} + T_{\rm sky} = 80 + 180 \big(\nu/{\rm 180\,MHz}\big)^{-2.6} \, {\rm K}$. The sky temperature ($T_{\rm sky}$) is driven from a relatively quiet zone on the sky \citep{Haslam1982} and the receiver temperature ($T_{\rm rec}$) is adopted from the mean receiver temperature studied by \citet{Razavi-Ghods2017}. The system temperature is used to construct the noise variance in the visibility space in Jy as follows:
\begin{align}
    \sigma_n = \frac{2k_B\nu^2\Omega_p}{c^2}\frac{T_{\rm sys}}{\sqrt{\Delta\nu \Delta t}},
\end{align}
where $\Omega_p$ is the solid angle of the primary beam. The noise visibilities are created by using Equation~\eqref{eqn:noise_sim}. Similar to the approach used for Phase~I observations, we simulate 200 random noise realizations assuming 300 night observations and transform them into images after applying \texttt{DAYENU}-filtering with the same filtering configuration of the 21~cm simulation data as described above. To remove the noise bias in $m_2$, we split the visibility data into even and odd time groups and create the maps separately, as we do for Phase I observations.

The intensity maps for the 21~cm fiducial model and noise simulations are generated over a 60$\times$10 square degree area by stitching together maps centered at each hour from 2 to 5 hrs, each map being 15$\times$10 square degrees in size. Since we simulate $EE$-polarization only, we convert the noise maps to pseudo stokes-I maps by multiplying $1/\sqrt{2}$, assuming the effect of each polarization on noise map is approximately symmetric and each polarization map is uncorrelated. This enhances the sensitivity of statistical measurements. The 21~cm maps remain the same, assuming $m_{EE} \approx m_{NN} \approx m_I$.

Since our wedge-filtering and mapmaking processes are linear, the 21~cm signal and noise components can be treated independently throughout the matrix transformation. Therefore, we construct noisy 21~cm maps by directly adding the separately processed 21 cm and noise maps, ensuring that correlation effects introduced by the mapmaking and filtering steps are fully accounted for. The $m_2$, $m_3$ and $S_3$ are then measured from the noisy image cubes that are binned into 2~MHz sliced volume.

In Section~\ref{sec:instrument_effects_1pnt_stat}, we analyzed the role of sample variance in measuring variance and skewness for Phase~I observations. To systematically assess the contributions of different uncertainty components in Phase~II observations, we first evaluate the detectability of $m_2$, $m_3$, and $S_3$ considering only sample variance. We then extend this analysis to include both sample variance and thermal noise to provide a more practical forecast.

The left panels of Figure~\ref{fig:forecast_1pnt_stat_h320} present the results for $m_2$ (top), $m_3$ (middle), and $S_3$ (bottom) from the HERA-320 simulations, assuming no foregrounds (i.e., perfect foreground removal without any loss of the 21 cm signal). The dashed lines represent the mean values averaged across different samples to estimate sample variance, without considering thermal noise, while the dark shaded regions indicate the 1$\sigma$ errors associated with sample variance. As discussed in Section~\ref{sec:instrument_effects_1pnt_stat}, there are relatively larger uncertainties in $m_3$ where sign transitions occur, while $m_2$ has tight sample variance across frequencies.

When thermal noise is included, the mean trends averaged over different samples and thermal noise realizations, shown by solid lines, remain similar to those of the dashed lines but with increased uncertainties. For $m_3$, the light shaded regions, representing 1$\sigma$ errors that account for both thermal noise and sample variance, become particularly significant at $z\sim8.5$--$9.5$, where sample variance is large. This may be explained by the extra sign-flip-instability introduced by the thermal noise. Additional substantial uncertainties appear at lower redshifts for all statistical quantities due to the receiver temperature, which becomes significant in $T_{\rm sys}$. The $S_3$ results suggest that there is potential to detect non-Gaussianities at 120--130~MHz and 160--170~MHz in the absence of foregrounds. For reference, the midpoint of reionization in this model occurs at $z \approx 8.5$, and at $z = 7.3$, where the fractional errors become large, the ionization fraction is 0.95.

\begin{figure}[t!]
\centering
\includegraphics[scale=0.35]{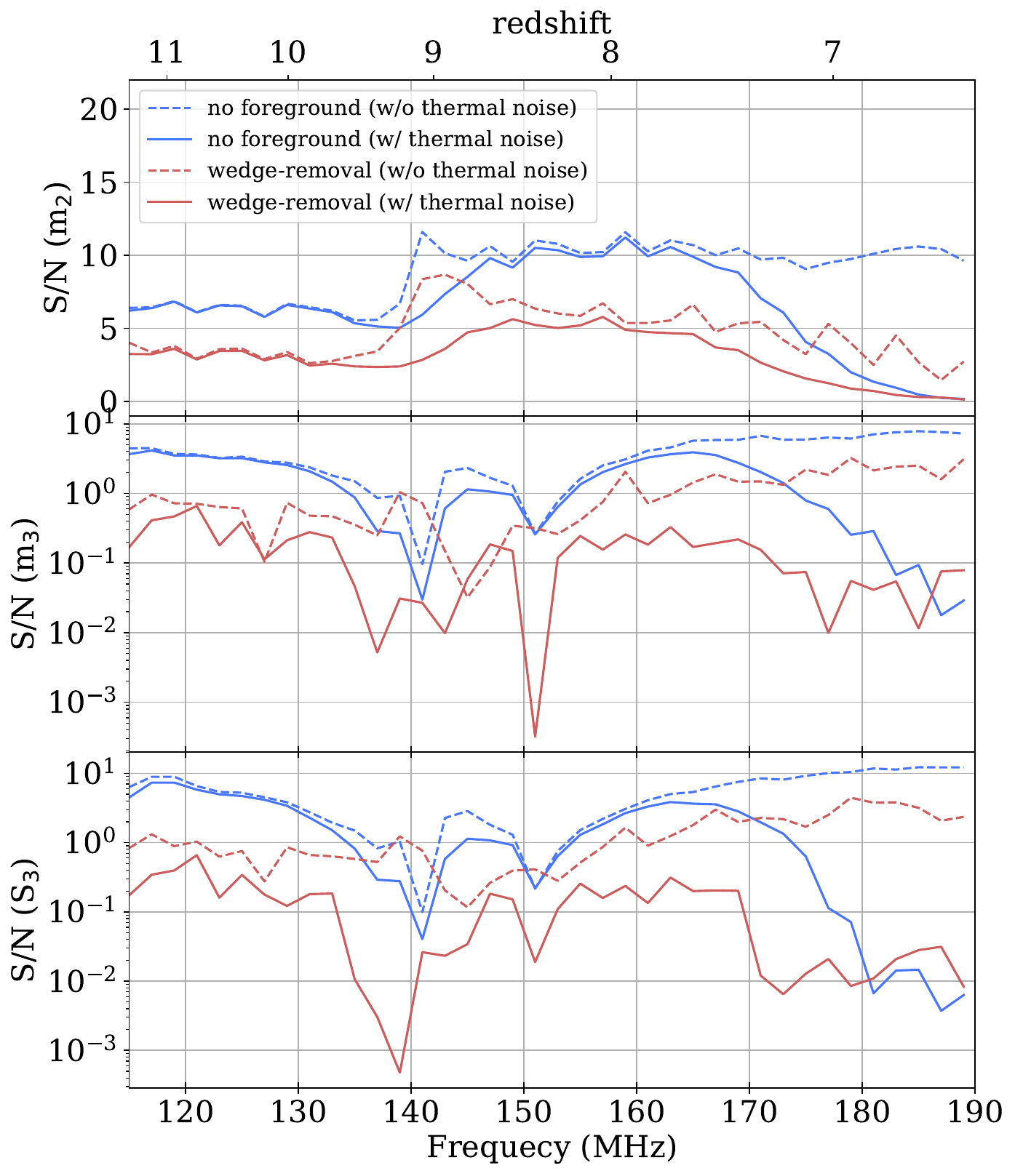}
\caption{Predictions of S/N for $m_2$ (top), $m_3$ (middle), and $S_3$ (bottom) measurements assuming 300 night observations, as forecasted with HERA-320 using the fiducial EoR model. We present the S/N considering different foreground removal methods including no foreground (i.e., perfect foreground removal) and wedge removal, represented by blue and red lines, respectively. In addition, we illustrate the case when considering the sample variance only (dashed lines) and the thermal noise and sample variance added in quadrature (solid lines). For the case when accounting for thermal noise and wedge-filtering (red solid line), the total S/N of $m_3$ in Band~1 and Band~2 are around 1.0 and 0.6, respectively.}
\label{fig:SNR_H320}
\end{figure}

In Section~\ref{sec:instrument_effects_1pnt_stat}, we observed that wedge-filtering significantly reduces the magnitude of $m_3$ while increasing sample variance, making it difficult to clearly distinguish $m_3$ values from zero at $z > 7.5$. However, there remains a possibility of detecting non-zero $m_3$ at lower redshifts in the absence of receiver noise, where a significant portion of neutral hydrogen has been reionized. This is illustrated by the dashed lines and dark shaded regions in the right panels of Figure~\ref{fig:forecast_1pnt_stat_h320}.

The low-redshift regions are influenced by receiver temperature, leading to corruption by thermal noise. The right column of Figure~\ref{fig:forecast_1pnt_stat_h320} demonstrates that $S_3$ measurements are impacted by the thermal noise combined with sample variance (solid lines and light shaded regions) across redshifts. Particularly, the uncertainty blows up at low redshifts, reducing the likelihood of detecting non-Gaussianity from $m_3$ and $S_3$ at these redshifts when wedge-filtering is applied.

We investigate the detectability of $m_2$, $m_3$, and $S_3$ in terms of S/N. The magnitudes of mean values of $m_2$ and $m_3$ across different samples and noise realizations are used as our signal values in computing S/N. We estimate the S/N for two different foreground removal strategies: perfect foreground removal (i.e., no foreground) and wedge removal.

Figure~\ref{fig:SNR_H320} illustrates the predicted S/N of $m_2$ (top), $m_3$ (middle), and $S_3$ (bottom) for 300 night observations. The no foreground and wedge removal methods are presented by blue and red lines, respectively. In computing the noise, we account for scenarios with sample variance only (dashed lines) and both thermal noise and sample variance (solid lines).

If sample variance is the only source of noise, the S/N for $m_2$ measurements reaches around 5--10, assuming no foregrounds. When thermal noise is included alongside sample variance, the S/N remains similar, except for $z < 7.5$, where the S/N declines due to the impact of thermal noise. With wedge-filtering applied, the S/N drops to around 3--8 for sample variance alone and falls below 5 when accounting for additional thermal noise, especially at higher frequencies.

The $m_3$ and $S_3$ measurements exhibit a lower S/N compared to $m_2$. In an optimistic scenario, considering only sample variance without foregrounds, the S/N averages around 4 across frequencies. However, this decreases at higher frequencies when thermal noise is included, reducing the S/N to below one. After applying wedge-removal, the S/N hovers around one for the no-foreground case and drops to less than one with wedge-filtering, making it challenging to probe non-Gaussianity with high sensitivity. This could be enhanced by extending the observation period and incorporating outriggers in the antenna array.

Although the S/N is low when accounting for thermal noise and wedge-filtering (solid red curve), combining information from multiple frequency bins of $m_2$ and $m_3$, such as through the Fisher information matrix \citep{Fisher1935}, could still offer insights into exploring model parameters. For example, the S/N of $m_3$ estimated from combined data points in Band~1 and Band~2 are around 1.0 and 0.6, respectively, for the solid red curve. We defer the detailed examination of model parameters reconstructed from $m_2$ and $m_3$ measurements to future work.

Since wedge-filtering hampers the detection of non-Gaussian features by reducing signal-to-noise, minimizing its impact is essential. One approach is to reduce the wedge size using image-based power spectra rather than delay spectra \citep{Morales2018}, or by optimizing array configurations \citep{Murray2018}. A smaller wedge may enhance sensitivity to non-Gaussianity in 21~cm signals. Efforts are also ongoing to reduce signal loss by separating foregrounds from the signal using Gaussian Process Regression (GPR) \citep{Kern2020c, Mertens2020, Soares2021}. More recently, machine learning methods have been explored to recover 21~cm signals within the wedge after foreground removal \citep[e.g.,][]{Gagnon-Hartman2021, Acharya2024}. These advances offer promising prospects for future HERA observations.

\section{Summary}
\label{sec:summary}
In this study, we examined the properties of images constructed by the HERA radio interferometer, after removing bright foreground emissions. Using HERA Phase I observations, previously employed for power spectrum analyses (\Hthree{}), we measured one-point statistics. We go beyond previous studies by forward-modeling the observational data, 21 cm simulations, and noise simulations to account for pixel correlation effects caused by the instrument response and frequency correlations arising from the foreground removal technique.

The observational data includes 94 nights of observations for Band 1 and Band 2. We analyzed the data before and after systematics mitigation to study the impact of systematics on one-point statistics such as the second ($m_2$) and third ($m_3$) moments. The radio interferometric images were constructed using the direct optimal mapping method, which optimally down-weights visibility data with large noise variance. We mapped four different fields of the sky. Foregrounds were removed using the \texttt{DAYENU} wedge-filtering technique.

In Section~\ref{sec:instrument_effects_1pnt_stat}, we examined the effects of the
instrument PSF and wedge-filtering on one point statistics using simulations without
thermal noise. The PSF attenuates small-scale structure, significantly reducing the amplitudes of $m_2$ and $m_3$. Wedge-filtering further suppresses large-scale modes, driving $m_3$ and $S_3$ close to zero, with the exception of weak signals at low redshifts. We also examined the predicted one-point statistics for HERA-320, which utilizes 320 antennas, assuming no thermal noise. The results indicate that, by capturing additional small-scale information, HERA-320 offers a better chance to detect non-Gaussian features in $m_3$ and $S_3$ across redshifts, with relatively small sample variance. However, wedge-filtering continues to cause substantial signal loss, rendering the measurements indistinguishable from the zero baseline due to the sample variance. Despite this, $S_3$ shows a distinct rising trend toward lower redshifts.

Section~\ref{sec:cc_mitigation} presents a comparison of one-point statistics in observational data before and after systematics mitigation. We find that residual systematics introduce excess variance ($m_2$) and spurious non-Gaussianity into the wedge-filtered maps. While systematics mitigation reduces this effect, evidence of residual contamination remains, as shown by features in the maps (Figure~\ref{fig:hera_stripe_wedge_filtered}) and outliers in $m_3$ relative to noise expectations.

In Section~\ref{sec:model_comparison}, we compared observational one-point statistics with simulations. Among the four EoR models considered, the cold reionization model, characterized by higher variance, is disfavored in Band~2, consistent with prior power spectrum constraints (\Hb{}, \Hthree{}). However, the large synthesized beam size in Phase~I and signal suppression from wedge-filtering prevent the detection of meaningful $m_3$ signatures from the 21~cm signal.

Finally, we also performed forecasts for the full HERA core (HERA-320). Assuming perfect foreground removal and 300 nights of observation, deviations in $m_3$ and $S_3$ from Gaussianity are predicted at z $>$ 7.5. However, thermal noise significantly reduces sensitivity, especially at high frequencies. With wedge-filtering applied, detecting non-Gaussian features becomes even more difficult, as $m_3$ amplitudes fall below the noise floor.

We quantified the signal-to-noise ratio (S/N) of $m_2$, $m_3$, and $S_3$ under both ideal and wedge-filtered conditions. In the perfect foreground removal case, $m_2$ achieves S/N of 5-–10, but drops with frequency as thermal noise increases. Wedge-filtering reduces $m_2$ S/N below 5. For $m_3$ and $S_3$, the S/N averages around 4 in the ideal case, but falls below 1 with wedge-filtering. Still, combining statistics across frequencies may improve overall constraints, and higher-order moments like $m_3$ and $S_3$, which probe non-Gaussian features of ionized bubble distributions, remain potentially valuable.

Ultimately, wedge-filtering significantly diminishes non-Gaussian signals, highlighting the need for developing techniques that enable effective analysis within the wedge. 
Understanding the extent to which the instrumental effects of PSF and wedge-filtering diminish the $m_2$ and $m_3$ signals is model-dependent, as different models have structures on different scales.

\section*{Acknowledgements}
This material is based upon work supported by the National Science Foundation under Grant Nos. 1636646 and 1836019 and institutional support from the HERA collaboration partners. This research is funded in part by the Gordon and Betty Moore Foundation through Grant GBMF5212 to the Massachusetts Institute of Technology. Nicholas Kern acknowledges support from NASA through the NASA Hubble Fellowship grant \#HST-HF2-51533.001-A awarded by the Space Telescope Science Institute, which is operated by the Association of Universities for Research in Astronomy, Incorporated, under NASA contract NAS5-26555.

\bibliography{main}{}
\bibliographystyle{aasjournal}

\end{document}